\pdfoutput=1
\documentclass[prodmode,acmcsur]{acmsmall} 

\usepackage[ruled]{algorithm2e}

\SetAlFnt{\small}
\SetAlCapFnt{\small}
\SetAlCapNameFnt{\small}
\SetAlCapHSkip{0pt}
\IncMargin{-\parindent}
\usepackage[english]{babel}
\usepackage{pifont}
\usepackage{algorithmic}
\usepackage{amsmath}
\usepackage{amssymb}
\usepackage{array}
\usepackage{booktabs}
\usepackage[small,labelfont=bf, textfont=it]{caption}
\usepackage[usenames,dvipsnames]{colortbl}
\usepackage{dsfont}
\usepackage{pifont}
\usepackage{epsfig}
\usepackage{fmtcount}
\usepackage{footnote}
\usepackage{graphicx}
\usepackage{microtype}
\usepackage{multirow}
\usepackage{placeins}
\usepackage{rotating}
\usepackage{url}
\usepackage{verbatim}
\usepackage{xspace}
\usepackage[dvipsnames]{xcolor}
\usepackage{tikz}
\usepackage{xkeyval}
\usetikzlibrary{shapes,decorations,shadows}
\usepackage{boiboites}
\usepackage{mathtools, cuted}
\usepackage{pdflscape}
\usepackage[draft]{hyperref}
\usepackage{natbib}
\usepackage{wrapfig}
\usepackage{mathtools}
\usepackage{subcaption}

\usepackage{afterpage}

\definecolor{Gray}{gray}{0.9}

\usepackage{color}
\def\A{\ensuremath{\mathbf A}\xspace}

\newcommand{\new}[1]{{\textcolor{blue}{#1}}}
\renewcommand{\new}[1]{{#1}}
\newcommand{\newnew}[1]{{\textcolor{black}{#1}}}

\newcommand{\hide}[1]{}

\newcommand{\cross}{\textcolor{red}{\ding{56}}}
\newcommand{\tick}{\textcolor{OliveGreen}{\ding{52}}}

\newcommand{\slashburn}{\textsc{SlashBurn}\xspace}

\newcommand{\metis}{\textsc{METIS}\xspace}

\newcommand{\vog}{\textsc{VoG}\xspace}

\newcommand{\gbase}{\textsc{Gbase}\xspace}

\newcommand{\coarsenet}{\textsc{coarseNet}\xspace}
\newcommand{\netcondense}{\textsc{NetCondense}\xspace}

\newcommand{\perseus}{\textsc{Perseus}\xspace}
\newcommand{\timecrunch}{\textsc{TimeCrunch}\xspace}

\newcommand{\G}{G}
\newcommand{\V}{\mathcal{V}}

\newcommand{\Wikipedia}{{\tt Wikipedia}\xspace}

\newcommand{\Wikipediacontro}{{\tt Controversy}\xspace}

     \newcounter{intuition}
     
     \newsavebox{\coloredbgbox}
     
     \newcounter{observation}
     
   \newcounter{problem}
       
       \newenvironment{problem}{
       	\begin{lrbox}{\coloredbgbox}
       		\begin{minipage}{\dimexpr1\linewidth-2\fboxsep-2\fboxrule\relax}
       			\refstepcounter{problem}
       			\textbf{\textsc{Problem} \theproblem.}
       		}
         		{
       		\end{minipage}
       	\end{lrbox}%
       	\begin{center}
       		\fcolorbox{black}{black!06}{\usebox{\coloredbgbox}}
       	\end{center}
       }

\acmVolume{V}
\acmNumber{N}
\acmArticle{A}
\acmYear{2017}
\acmMonth{0}

\begin{document}
\markboth{Y. Liu, T. Safavi, A. Dighe and D. Koutra}{Graph Summarization Methods and Applications: A Survey}

\title{Graph Summarization Methods and Applications: A Survey}
\author{YIKE LIU
\affil{University of Michigan, Ann Arbor}
TARA SAFAVI
\affil{University of Michigan, Ann Arbor}
ABHILASH DIGHE
\affil{University of Michigan, Ann Arbor}
DANAI KOUTRA
\affil{University of Michigan, Ann Arbor}
}

\begin{abstract}
While advances in computing resources have made processing enormous amounts of data possible, human ability to identify patterns in such data has not scaled accordingly. Efficient computational methods for condensing and simplifying data are thus becoming vital for extracting actionable insights. In particular, while data summarization techniques have been studied extensively, only recently has summarizing interconnected data, or \emph{graphs}, become popular. 
This survey is a structured, comprehensive overview of the state-of-the-art methods for summarizing graph data. We first broach the motivation behind, and the challenges of, graph summarization. We then categorize summarization approaches by the type of graphs taken as input and further organize each category by core methodology. Finally, we discuss applications of summarization on real-world graphs and conclude by describing some open problems in the field.
\end{abstract}

\maketitle

\section{Introduction}
\label{sec:intro}
As technology advances, the amount of data that we generate and our ability to collect and archive such data both 
increase continuously.
Daily activities like social media interaction, web browsing, product and service purchases, itineraries, and wellness sensors generate large amounts of data, the analysis of which can immediately impact 
our lives. This abundance of generated data and its velocity call for data summarization, one of the main data mining tasks. 

Since summarization facilitates the identification of structure and meaning in data, the data mining community has taken a strong interest in the task.
Methods for a variety of data types have been proposed:
sequence data and events~\cite{garriga:05:summarizing}, 
itemsets and association rules~\cite{Liu1999,yan:05:profiles,Ordonez2006b,mampaey:11:techrepkdd}, 
spatial data~\cite{Lin2003},
transactions and multi-modal databases~\cite{wang:04:sumtrans,chandola:05:summ,Shneiderman08,xiang:10:hyper}, 
data streams and time series~\cite{CM05,palpanas:08:timeseriessum},
video and surveillance data~\cite{DBLP:conf/icdm/PanYF04,conf/wiamis/DamnjanovicAI008}, and
activity on social networks~\cite{LinSK08,Mehmood2013}. 

This survey focuses on the summarization of interconnected data, otherwise known as graphs or networks, a problem in graph mining with connections to relational data management and 
visualization. Graphs are ubiquitous, representing a variety of natural processes as diverse as friendships between people, communication patterns, and interactions between neurons in the brain. 
Formally, 
a \textit{plain} graph or network is an abstract data type consisting of a finite set of vertices (nodes) $\V$ and a set of links (edges) $\mathcal{E}$.
The latter represent interactions between pairs of vertices. 
A graph is often represented by its adjacency matrix \A, which can be binary, corresponding to whether there exists an interaction between two vertices, or numerical, corresponding to the strength of the connection. We will refer 
to a graph with numerical or categorical labels (attributes or annotations) for its nodes or edges as a \textit{labeled} graph. A network that changes over time is called \textit{dynamic} or \textit{time-evolving} and is often described by a series of adjacency matrices, one per timestamp. Examples of graphs are social networks, traffic networks, computer networks, phone call or messaging networks, location check-in networks, protein-protein interaction networks, user-product review or purchase networks, functional or structural brain connectomes, and more.

The benefits of graph summarization include: 
\vspace{-0.2cm}
\begin{itemize}
    \item \emph{Reduction of data volume and storage.} Graphs of real-world datasets are often massive. For example, \new{as of August 2017 the Facebook social network had 2 billion users}, and more than 100 billion emails were exchanged daily. Summarization techniques produce small summaries that require significantly less storage space than their original counterparts. Graph summarization techniques can decrease the number of I/O operations, reduce communication volume between clusters in a distributed setting, allow loading the summary graph into memory, \new{and facilitate the use of graph visualization tools  while avoiding the ``hairball'' visualization problem}. 
    \item \emph{Speedup of graph algorithms and queries.} While a plethora of graph analysis methods exist, many cannot efficiently handle large graphs. Summarization techniques produce smaller graphs that maintain the most salient information from the original graph. The resultant summary graph can be queried, analyzed, and understood more efficiently with existing tools and algorithms.
    \item \emph{Interactive analysis support.} As the systems side makes advancements in interactive graph analysis, summarization is introduced to handle information extraction and speed up user analysis. The resultant graph summaries make it possible to visualize datasets that are originally too large to load into memory.
    \item \emph{Noise elimination.} Real graph data are frequently large-scale and considerably noisy with many hidden, unobserved, or erroneous links and labels.
    Such noise hinders analysis by increasing the workload of data processing and hiding the more ``important" information. Summarization serves to filter out noise and reveal patterns in the data. 
\end{itemize}

\vspace{-0.2cm}
Given its advantages, graph summarization has extensive applications, including 
clustering~\cite{cilibrasi:05:cluster},  classification~\cite{leeuwen:06:class}, community detection ~\cite{ChakrabartiPMF04}, 
outlier detection~\cite{smets:11:odd,akoglu:12:comprex}, pattern set mining~\cite{vreeken:11:krimp}, 
finding sources of infection in large graphs~\cite{prakash:12:netsleuth}, 
and more.

\new{The problem of graph summarization has been studied algorithmically in the fields of graph mining and data management, 
while interactive exploration of the data and appropriate display layouts have been studied in visualization. In this survey, we review graph summarization mostly from a methodological perspective, answering how we can algorithmically obtain summaries of graph data.
We also give pointers to visual analytics platforms that can consume algorithmic outputs and explore display options. } 

\enlargethispage{2\baselineskip}
\subsection{Challenges}
Overall, the notion of a graph summary is not well-defined. 
A summary is application-dependent and can be defined with respect to various goals: it can preserve specific structural patterns, focus on some network entities, preserve the answers to graph queries, or maintain the distributions of graph properties. Overall, graph summarization has five main challenges: 

\begin{enumerate}
    \item \emph{Data volume.} The main target of graph summarization is to reduce the size of the input graph data so that other analyses can be performed efficiently. At the same time, though, summarization techniques are themselves faced with the challenge of processing large amounts of data. The requirement of efficiency often steers their design 
    toward techniques that scale well with the size of the input graph. \new{Table~\ref{tab:static_qualitative} points to methods that are linear on the size of the input.} 
    
    \item \emph{Complexity of data.} Graph operations often cannot be easily partitioned and parallelized because of the many interactions between entities, as well as the complexity of entities themselves. Furthermore, the heterogeneity of nodes and edges continues to increase in real networks.
    Accordingly, incorporating side information from heterogeneous sources (text, images, etc.) may require highly detailed design and quantification in algorithms.
    For example, in social networks, users can chat or share with each other, follow or friend each other, and a single user profile alone contains a great deal of information. Finally, real datasets often contain noise or missing information, which may interfere with the pattern mining process. 
    new{Sections~\ref{sec:attribute} and~\ref{sec:dynamic} review methods for attributed and dynamic networks, which tend to be more complex than methods for plain networks.} 
    \item \emph{Definition of interestingness.} Summarization involves extracting of important or interesting information. However, the definition of ``interesting'' is itself subjective, usually requiring both domain knowledge and user preferences. Moreover, the cutoff between ``interesting'' and ``uninteresting'' can be difficult to determine in a principled way; usually it is decided by considering the tradeoffs between time, space, and information preserved in the summary, as well as the complexity of mapping solutions obtained from the summary back onto the original nodes and edges. \new{Each presented graph summarization technique uses different optimization formulations to define the interestingness of a summary.}
    \item \emph{Evaluation.} \new{Evaluation of summarization outputs depends on the application domain. From the database perspective, a summary is good if it efficiently supports both global and local queries with high accuracy. 
    In the context of summarizing community information, either community preservation is maximized or reconstruction error is minimized. 
    Compression-based techniques seek to minimize the number of bits needed to describe the input graph, or else the number of nodes/edges or the normalized number of bits per edge. Furthermore, evaluations become more complex when more elements, such as visualization and multi-resolution summaries, are involved. In these cases, user studies and qualitative criteria may be employed.
    }
    \item \emph{Change over time.} Ideally, graph summaries should evolve over time, since real data are usually dynamic. For instance, social network activity, brain functions, and email communications---all naturally represented as graphs---change with time. How to incorporate the dynamic nature of such data in computation and perform analysis efficiently becomes an essential question. \new{Section~\ref{sec:dynamic} reviews methods that treat dynamic graphs as a sequence of static snapshots or streams.} 
\end{enumerate}

\vspace{-0.2cm}
As demonstrated by these challenges, graph summarization is a difficult and multifaceted problem.

\enlargethispage{2\baselineskip}
\subsection{\new{Types of Graph Summaries}}

\new{
In this survey, we categorize graph summarization methods based on the type of data handled and the core techniques employed.
Below we give the main types of graph summaries.
Table~\ref{tab:static_qualitative} provides detailed information for each approach.
}

\new{$\bullet$ \textbf{Input: Static or Dynamic.} Most summarization methods operate on static networks, leveraging graph structure (links) and, if available, the node/edge attributes. Despite the prevalence of large dynamic networks, only recent research efforts address their efficient summarization. In some cases, static methods are adapted to handle dynamic networks seen as series of static snapshots.
In other cases, new methods for graph streams are devised. In this survey, we first categorize summarization methods based on their input type (Figure~\ref{fig:taxonomy}).}

\new{$\bullet$ \textbf{Input: Homogeneous or Heterogeneous.} The most well-studied instance in  graph summarization, and graph mining more generally, is the homogeneous graph with one entity and one link type.
However, some approaches apply to heterogeneous graphs by treating various types of nodes (e.g., students, instructors) and relations between them (e.g., teacher, friends, classmates) differently. These methods tend to be more complex, but also more expressive.}

\new{$\bullet$ \textbf{Core Technique.} Across the literature, graph summarization methods employ a set of core techniques:} 
\vspace{-0.2cm}
\new{\begin{itemize}
    \item \textbf{Grouping- or aggregation-}based: This is the most popular technique. Some \textit{node-grouping} methods recursively aggregate nodes into ``supernodes'' based on an application-dependent optimization function, which can be based on structure and/or  attributes.
    Others employ existing clustering techniques and map each densely-connected cluster to a supernode. \textit{Edge-grouping} methods aggregate edges into compressor or virtual nodes. 
    \item \textbf{Bit compression-}based: This approach, a common technique in data summarization, minimizes the  number of bits needed to describe the input graph via its summary. 
    Some methods are lossless and can perfectly reconstruct the original graph from the summary. Others are lossy, compromising recovery accuracy for space savings.  
    \item \textbf{Simplification- or sparsification-}based: These methods streamline an input graph by removing less ``important'' nodes or edges, resulting in a sparsified graph. 
    \item \textbf{Influence-}based: These approaches aim to discover \newnew{a high-level description of the  influence propagation} in large-scale graphs.
    Techniques in this category formulate the summarization problem as an optimization process in which some quantity related to information influence is maintained.
\end{itemize}
}
\new{$\bullet$ \textbf{Output: Summary Type.} The output of a summarization approach can be: (i) a supergraph, which consists of supernodes or collections of original nodes, and superedges between them; (ii) a sparsified graph, which has fewer nodes and/or edges than the original network; 
or (iii) \newnew{a list of (static or temporal) structures or influence propagations, which are seen independently instead of in the form of a single summary graph}. 
Moreover, the summary can be: (a) flat, with nodes simply grouped into supernodes, or (b) hierarchical, with multiple levels of abstraction.}

\new{$\bullet$ \textbf{Output: Non-overlapping or Overlapping Nodes.} In its simplest form, a summary is non-overlapping: each original node belongs only to one summary element (e.g., supernode, subgraph). Overlapping summaries, where a node may belong to multiple elements, can  capture complex inherent data relationships, but may also complicate intepretation and visualization.}

\new{$\bullet$ \textbf{Main Objective.} The key objectives of graph summarization include query efficiency and approximate computations, compression and data size reduction, static or temporal pattern discovery, visualization and interactive large-scale visual analytics, influence analysis and understanding, entity resolution, and privacy preservation.}

\subsection{\new{Differences from Prior Surveys}}

\begin{figure}[h!]
\centering

\vspace{-0.4cm}
\includegraphics[width=0.95\textwidth]{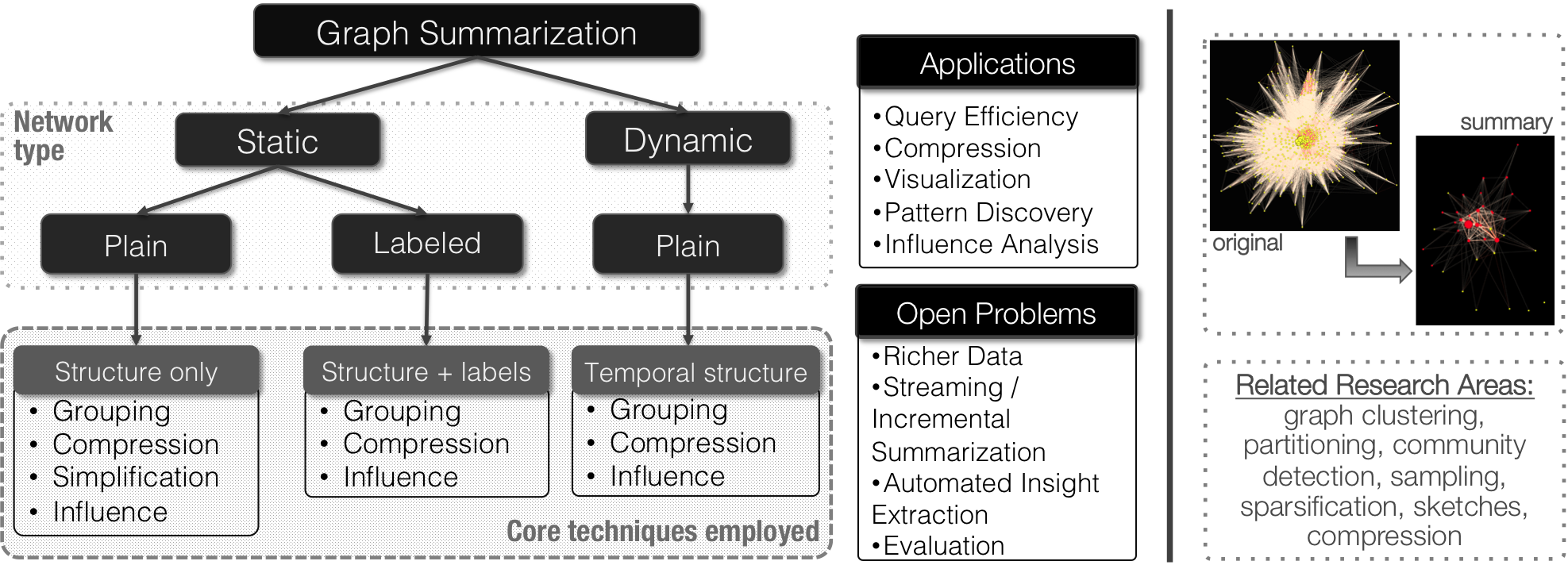}
\caption{\new{Overview of our survey}. Taxonomy of graph summarization algorithms based on the input type and the core employed technique; alternative approaches; applications; and open problems. 
}
\label{fig:taxonomy}
\vspace{-0.25cm}

\end{figure}

Previous work on surveying the graph summarization literature is scarce. \citet{you2013towards} present some summarization algorithms for \emph{static} graphs, focusing mostly on grouping- and compression-based methods. 
The tutorial by ~\citeauthor{tutorial}~\citeyear{tutorial} provides more specific categorization and descriptions of ongoing work, but again only addresses static graph summarization. 
By contrast, we review a wide set of proposed methodologies for both static \emph{and} dynamic graph summarization.
Specifically, in this survey: 
\vspace{-0.3cm}
\begin{enumerate}
    \item We create a taxonomy (Figure~\ref{fig:taxonomy}) on \new{the three main instances of the graph summarization problem: for plain static graphs (Section~\ref{sec:static}), for static graphs with additional side information or labels (Section~\ref{sec:attribute}), and for (plain) graphs that evolve over time (Section~\ref{sec:dynamic}). Within each instance of the problem, we present key algorithmic ideas and methodologies used to solve it.}  
    \item We highlight methodological properties  that are  useful to researchers and practitioners, such as input/output data types and end goal (for example, compression vs.\ visualization), \new{and present them concisely in Table~\ref{tab:static_qualitative}.}
    \item We give connections between methods of graph summarization and related fields that, \newnew{while not directly supporting graph summarization}, have potential in summarization tasks. These fields include compression, sparsification, and clustering and community detection.
    \item We review real-world applications of graph summarization and identify 
    open problems and opportunities for future research (Sections~\ref{sec:application} and~\ref{sec:conclusion}).
\end{enumerate}





\tikzset{
  basic/.style  = {draw, text width=1cm, drop shadow, font=\sffamily\small, rectangle},
  root/.style   = {basic, rounded corners=2pt, thin, align=center,
                   fill=blue!20!black, text width=10em, text=white},
  level 2/.style = {basic, rounded corners=2pt, thin,align=center, fill=blue!50!black,
                   text width=8em, text=white, blur shadow={shadow blur steps=10}},
  level 3/.style = {basic, rounded corners=2pt, thin,align=center, fill=blue!70!black,
                   text width=9em, text=white, blur shadow={shadow blur steps=10}},
  level 4/.style = {basic, thin, align=left, fill=white!95!black, text width=6.5em, text=black, blur shadow={shadow blur steps=5}}
}

\section{Static Graph Summarization: Plain Networks}
\label{sec:static}
\label{sec:structure}

\enlargethispage{3\baselineskip}
Most work in static graph summarization focuses solely on graph structure \newnew{without side information or labels}. 
At a high level, the problem of summarization or aggregation or coarsening of static, plain graphs is described as:

\begin{problem} \textbf{Summarization of Static, Plain Graphs}. \hfill \\
\textbf{Given} a static graph $\G$ without any side information, or its adjacency matrix $\A$, \\
{\bf Find} a summary graph or a set of structures or a compressed data structure to concisely describe the given graph.
\end{problem}

The first block of Table~\ref{tab:static_qualitative} 
qualitatively compares and explicitly characterizes static graph summarization methods for plain networks. 
Here, we review these methods by organizing them into categories based on the core methodology that they employ for the summarization task. When applicable, we first give the high-level idea per method type and then describe the corresponding technical details.

\subsection{\bf Grouping-based methods}

\new{Grouping-based methods are among the most popular techniques for summarization. We distinguish grouping-based graph summarization methods into two main categories: (i) node-grouping and (ii) edge-grouping. 
In Section~\ref{sec:compression}, we discuss methods that use bit-level compression as their primary summarization technique and grouping as a complementary technique.}

\subsubsection{\bf Node-grouping methods}
\label{subsec:node-grouping}
\textit{Some approaches employ existing clustering techniques to find clusters that then map to supernodes. Others recursively aggregate nodes into supernodes, connected via superedges, based on an application-dependent optimization function.} 

\emph{Node clustering-based methods}. 
\new{Although node grouping and clustering are related in that they result in collections of nodes, they have different goals. In the context of summarization, node grouping is performed so that the resultant graph summary has specific properties, \makebox[\textwidth][s]{e.g., query-specific properties or maintenance of  edge weights. On the other hand,}}

\begin{savenotes}
\begin{landscape}
\begin{table}[t!]
\centering
\scriptsize
\caption{Qualitative comparison of all graph summarization techniques based on the properties of the input graph (e.g., weighted, (un)directed, homogeneous/heterogeneous), their algorithmic properties (i.e., user-defined parameters, complexity linear on the number of edges, core technique, output), and their main objectives. Notation: $^*$ for the input means that the algorithm can be extended to that type of input, but details are not in the paper; for complexity $^*$ indicates sub-linearity.} 
\label{tab:static_qualitative}
\resizebox{1.45\textwidth}{!}{
\begin{tabular}{|c|p{3cm}|cccc|cccc|c|} \hline
& & \multicolumn{4}{|c|}{\textbf{Input Graph}} & \multicolumn{4}{|c|}{\textbf{Algorithmic Properties}} & \\ \hline
& \multicolumn{1}{|c|}{\textbf{Method}} &
\multicolumn{1}{c}{{ \begin{sideways}\textbf{Weighted} \end{sideways}}} &
\multicolumn{1}{c}{{ \begin{sideways}\textbf{Undirect.} \end{sideways}}} &
\multicolumn{1}{c}{{ \begin{sideways}\textbf{Directed} \end{sideways}}} &
\multicolumn{1}{c|}{{ \begin{sideways}\textbf{Heterog.} \end{sideways}}} &
\multicolumn{1}{c}{{ \begin{sideways}\textbf{Prm-free} \end{sideways}}} &
\multicolumn{1}{c}{{ \begin{sideways}\textbf{Linear} \end{sideways}}} &
\multicolumn{1}{c}{{\textbf{Technique}}} &
\multicolumn{1}{c|}{{ \textbf{Output}}} &
\multicolumn{1}{c|}{{ \textbf{Objective}}}\\
 \hline 
\rowcolor{Gray} \cellcolor{white!50}  &  \multicolumn{1}{p{5cm}|}{GraSS  {\scriptsize \cite{lefevre2010grass}}}
 & \cross & \tick & \cross & \cross & \cross & \cross & grouping & supergraph & query efficiency \\ 
 &  \multicolumn{1}{p{5cm}|}{Weighted Compr.  {\scriptsize \cite{ToivonenZHH11}}}
& \tick & \tick & \tick$^*$ & \cross & \cross & \cross & grouping &
supergraph & compression\\ 
\rowcolor{Gray} \cellcolor{white!50}  &  \multicolumn{1}{p{5cm}|}{\coarsenet   {\scriptsize \cite{purohit2014fast}}} 
& \tick & \cross & \tick & \cross & \cross & \tick$^*$ & 
grouping & supergraph & influence\\  
  &  \multicolumn{1}{p{5cm}|}{$l_p$-reconstr. Error  {\scriptsize \cite{riondato2014graph}}}
& \tick & \tick & \cross & \cross & \cross & \tick & 
grouping & supergraph & query efficiency\\  
\rowcolor{Gray} \cellcolor{white!50}  &  \multicolumn{1}{p{5cm}|}{Motifs   {\scriptsize \cite{DunneS13}}}
& \cross & \tick & \cross & \tick & \cross & \cross &  
grouping & supergraph & visualization \\ 
&  \multicolumn{1}{p{5cm}|}{CoSum   {\scriptsize \cite{zhu2016unsupervised}}}
& \tick$^*$\footnote{Weights only exist between nodes of the same type.} & \tick & \cross & \tick & \tick & \tick & grouping & supergraph & entity resolution\\ 
 \rowcolor{Gray} \cellcolor{white!50}  &  \multicolumn{1}{p{5cm}|}{Dedensification  {\scriptsize \citep{maccioni2016dedensification}}}
& \cross & \tick$^*$ & \tick & \tick & \tick & \tick$^*$ & (edge) grouping & sparsified graph & query efficiency \\  
&   \multicolumn{1}{p{5cm}|}{VNM {\scriptsize \cite{buehrer2008scalable}}}
& \cross & \cross & \tick & \cross & \cross & \cross & 
(edge) grouping & sparsified graph & patterns \\ 
\rowcolor{Gray} \cellcolor{white!50}  &  \multicolumn{1}{p{5cm}|}{MDL Repres.  {\scriptsize \cite{NavlakhaRS08}}}
& \cross & \tick & \tick$^*$ &\cross & \tick & \cross &  
compression & supergraph & compression\\ 
 &  \multicolumn{1}{p{5cm}|}{\vog  {\scriptsize \cite{koutra:14:vog}}}
 & \cross & \tick & \cross &  \cross & \tick & \tick$^*$ & 
compression & structure list & patterns, visualiz. \\  
\rowcolor{Gray} \cellcolor{white!50}  &   \multicolumn{1}{p{5cm}|}{OntoVis  {\scriptsize \cite{shen2006visual}}}
& \cross & \tick & \cross & \tick & \tick & \tick & 
simplification & sparsified graph & visualization \\ 
&  \multicolumn{1}{p{5cm}|}{Egocentric Abstr.  {\scriptsize \cite{li2009egocentric}}}
& \cross & \cross & \tick & \tick & \cross & \cross &  
simplification & sparsified graph & influence \\  
\rowcolor{Gray} \cellcolor{white!50}  &  \multicolumn{1}{p{5cm}|}{CSI  {\scriptsize \cite{Mehmood2013}}}
& \cross & \cross & \tick & \cross & \tick & \cross &
influence & supergraph & influence\\ 
 \multirow{-14}{*}{\rotatebox[origin=c]{90}{Static Plain Graphs}} \ &  \multicolumn{1}{p{5cm}|}{SPINE  {\scriptsize \cite{MathioudakisBCGU11}}}
 & \tick & \cross & \tick & \cross & \cross & \cross & influence & sparsified graph & influence\\  \midrule
 
\rowcolor{Gray} \cellcolor{white!50}  &  \multicolumn{1}{p{5cm}|}{S-Node  {\scriptsize \cite{raghavan2003representing}}}
 & \cross & \cross & \tick & \cross & \tick & \cross & grouping &   
supergraph  & 
query efficiency  \\  
&   \multicolumn{1}{p{5cm}|}{SNAP/k-SNAP  {\scriptsize \cite{tian2008efficient}}}
 & \cross & \tick & \tick$^*$ & \cross & \tick & \tick & grouping &  
supergraph &  
query efficiency  \\ 
\rowcolor{Gray} \cellcolor{white!50}  &  \multicolumn{1}{p{5cm}|}{CANAL  {\scriptsize \cite{ZhangTP10}}}
 & \cross & \tick & \tick$^*$ & \cross & \tick & \cross & grouping & 
supergraph  &
patterns  \\ 
 &  \multicolumn{1}{p{5cm}|}{Probabilistic  {\scriptsize \cite{hassanlou2013probabilistic}}}
 & \tick & \cross & \tick & \cross & \tick & \tick & grouping & supergraph & compression  \\ 
\rowcolor{Gray} \cellcolor{white!50}  &   \multicolumn{1}{p{5cm}|}{Query-Pres.  {\scriptsize \cite{fan2012query}}}
 & \cross & \cross & \tick &\cross & \cross & \tick  & grouping & supergraph & query efficiency\\  
&   \multicolumn{1}{p{5cm}|}{ZKP  {\scriptsize \cite{shoaran2013zero}}}
& \cross & \cross & \tick & \cross & \tick & \tick & grouping & supergraph & privacy  \\  
\rowcolor{Gray} \cellcolor{white!50}  &   \multicolumn{1}{p{5cm}|}{Randomized   {\scriptsize \cite{chen2009mining}}}
 & \cross & \tick & \cross & \tick & \tick & \cross & grouping & supergraph  & 
patterns  \\
&   \multicolumn{1}{p{5cm}|}{$d$-summaries  {\scriptsize \cite{song2016knowledgegraphs}}}
 & \cross & \cross & \tick & \cross & \cross & \cross & grouping & 
supergraph  & 
query efficiency  \\
\rowcolor{Gray} \cellcolor{white!50}  &  \multicolumn{1}{p{5cm}|}{SUBDUE  {\scriptsize \cite{cook:94:subdue}}}
 & \cross & \tick & \tick & \tick & \tick & \cross & 
compression & supergraph  & 
patterns  \\ 
&   \multicolumn{1}{p{5cm}|}{AGSUMMARY  {\scriptsize \cite{wu2014graph}}}
 & \cross & \cross & \tick & \cross & \tick & \tick & compression  & supergraph & compression  \\ 
 \rowcolor{Gray} \cellcolor{white!50}  &   \multicolumn{1}{p{5cm}|}{LSH-based  {\scriptsize \cite{khan2014set}}}
 & \cross & \cross & \tick & \cross & \cross & \tick & compression & supergraph  & compression  \\ 
 \multirow{-11}{*}{\rotatebox[origin=c]{90}{Static Labeled Graphs}} 
&   \multicolumn{1}{p{5cm}|}{VEGAS  {\scriptsize \cite{shi2015vegas}}}
 & \tick$^*$ & \cross & \tick & \cross & \cross & \tick$^*$ & influence & 
supergraph  & 
influence  \\  \midrule
\rowcolor{Gray} \cellcolor{white!50} &  \multicolumn{1}{p{5cm}|}{\netcondense~\scriptsize{\cite{adhikari2017condensing}}}  & \tick & \tick & \tick & \cross & \cross & \cross & grouping  & temporal supergraph & influence \\
 &   \multicolumn{1}{p{5cm}|}{TCM  {\scriptsize \cite{tang2016graph}}}
 & \tick & \tick & \tick & \cross & \cross & \tick & grouping  & supergraph & query efficiency  \\ 
\rowcolor{Gray} \cellcolor{white!50}   &   \multicolumn{1}{p{5cm}|}{}  &  &  &   &  & &   &    
  & ranked list of & temporal patterns,     \\ 
\rowcolor{Gray} \cellcolor{white!50}  &  \multicolumn{1}{p{5cm}|}{{\multirow{-2}{*}{TimeCrunch \scriptsize \cite{ShahKZGF15}}}}
 & \multirow{-2}{*}{\cross} & \multirow{-2}{*}{\tick} & \multirow{-2}{*}{\cross} &   \multirow{-2}{*}{\tick} & \multirow{-2}{*}{\cross}  & \multirow{-2}{*}{\tick} & \multirow{-2}*{compression} &  temporal structures & visualization   \\   
 & \multirow{2}{*}{\textsc{OSNet}~{\scriptsize \cite{QuLJZF14}}}
& \multirow{2}{*}{\cross} & \multirow{2}{*}{\tick} & \multirow{2}{*}{\cross} &  \multirow{2}{*}{\cross} & \multirow{2}{*}{\cross}  & \multirow{2}{*}{\tick} & \multirow{2}*{influence} & subgraphs of diffusion &  \multirow{2}{*}{influence}     \\  
  & \multicolumn{1}{p{5cm}|}{}
  &  &  &   &  &   & &  & over time &       \\  
\rowcolor{Gray} \cellcolor{white!50}  \multirow{-7}{*}{\rotatebox[origin=c]{90}{Dynamic Graphs}}  &  \multicolumn{1}{p{5cm}|}{Social Activity {\scriptsize \cite{LinSK08}}} & \cross & \tick & \cross &  \tick & \cross  & \cross & influence & temporal themes &  influence, visualization    \\  
  \hline
\end{tabular}
}
\end{table}
\end{landscape}
\end{savenotes}

\noindent \new{clustering or partitioning usually targets the minimization of cross-cluster edges or a variant thereof, without the end goal of producing a graph summary. Moreover, unlike role mining~\cite{conf/kdd/HendersonGLAETF11,conf/kdd/HendersonGETBAKFL12,GilpinRD13} or structural equivalence~\cite{PelegS89}, which seek to identify ``functions'' of nodes (e.g., bridge or spoke nodes) and find role memberships, summarization methods seek to group nodes that have not only structural similarities, but are also connected or close to each other in the network and thus can be replaced with a supernode. }

 \new{Although the goal of clustering is not graph summarization, the outputs of clustering algorithms can be easily converted to  non-application-specific summaries. 
 In a nutshell, a small representation of the input graph can be obtained by (i) mapping all the nodes that belong to the same cluster / community to a supernode, and (ii) linking them with superedges with weight equal to the sum of the cross-cluster edges, or else the sum of the weights of the original edges~\cite{NewmanG03,Yang:2013,Low:2012}. } \new{Although the clustering output can be viewed as a summary graph, a fundamental difference from tailored summarization techniques is that the latter groups nodes that are linked to the rest of the graph in a similar way, while clustering methods simply group densely-connected nodes. There exist comprehensive introductions to clustering techniques ~\cite{leskovec2014mining,aggarwal2015data} and work on clustering or community detection methods~\cite{aggarwal2010survey}, so we do not cover them in this survey. Among the most popular partitioning methods are Graclus~\cite{graclus}, spectral partitioning~\cite{alpert1999spectral}, and METIS~\cite{Karypis99@METIS}.} \newnew{Although METIS is a well-known partitioning approach that finds ``hard'' node memberships, it 
constructs a series of graph ``summaries'' by iteratively finding the maximal graph matching and merging nodes that are incident to an edge of the matching. 
The bisection result on the most coarsened graph is then projected backwards to the original graph. Via this process, it is possible to obtain a compact, hierarchical representation of the original graph, which resembles other node-grouping summarization methods. }

\vspace{0.1cm}
\emph{Node aggregation-based methods}.
One representative algorithm of hierarchical clustering-based node grouping is GraSS~\cite{lefevre2010grass}, which targets accurate query handling.
This summarization method supports queries on the adjacency between two nodes, as well as the degree and the eigenvector centrality of a node. The graph summaries are generated by greedily grouping nodes such that the normalized reconstructed error,  
$
\frac{1}{|\mathcal{V}|^2}  \sum_{i \in \mathcal{V}}\sum_{j \in \mathcal{V}}{|\bar{A}(i,j)-A(i,j)|}, $ is minimized---\new{$\mathbf{A}$ is the original adjacency matrix of the graph and $\mathbf{\bar{A}}$ is the real-valued approximate adjacency matrix, each entry of which intuitively represents the probability of the corresponding edge existing in the original graph given the summary. }
The resulting summaries are represented as a group of vertex sets with information about the number of edges within and between clusters. These sets are used to generate a probabilistic approximate adjacency matrix upon which incoming queries are computed. For example, if many edges cross vertex sets \emph{A} and \emph{B}, then it is likely that a node in \emph{A} is connected to a node in \emph{B}. In 
another variant, GraSS leverages Minimum Description Length (MDL) to automatically find the optimal number of supernodes in the summary. 

While GraSS does not guarantee output quality, 
~\citet{riondato2014graph} propose a method of generating supernodes and superedges with guarantees. 
Here, the objective is to find a supergraph that minimizes the $l_p$-reconstruction error, or the $p$-norm of  $\mathbf{A} - \mathbf{\bar{A}}$, as opposed to the normalized reconstruction error in GraSS, given a number of supernodes $k$. 
The proposed approach, which uses sketching, sampling, and approximate partitioning, is the first polynomial-time approximation algorithm of its kind with runtime $O(|\mathcal{E}| + |\mathcal{V}|\cdot k)$. 
This method targets efficiency for the same types of  queries as GraSS, as well as triangle and subgraph counting queries.

\new{~\citet{ToivonenZHH11} focus on compressing graphs with edge weights, proposing to merge 
nodes with similar relationships to other entities (structurally equivalent nodes) such that approximation error is minimized and  compression is maximized.} 
In merging nodes to obtain a compressed graph, the algorithm maintains either edge weights or strengths of connections of up to a certain number of hops. Specifically, 
in the simplest version of the solution, each superedge is assigned the mean weight of all edges it represents.  
In the generalized version, 
the best path between any two nodes is ``approximately equally good'' in the compressed graph and original graphs, but the paths do not have to be the same. \new{The definition of path ``goodness'' is data- and application-dependent. For example, 
the path quality can be defined as the maximum flow through the path for a flow graph, or the probability that the path exists for a probabilistic or uncertain graph.}
 
 \begin{figure}[t] 
\centering
\includegraphics[width=0.85\textwidth]{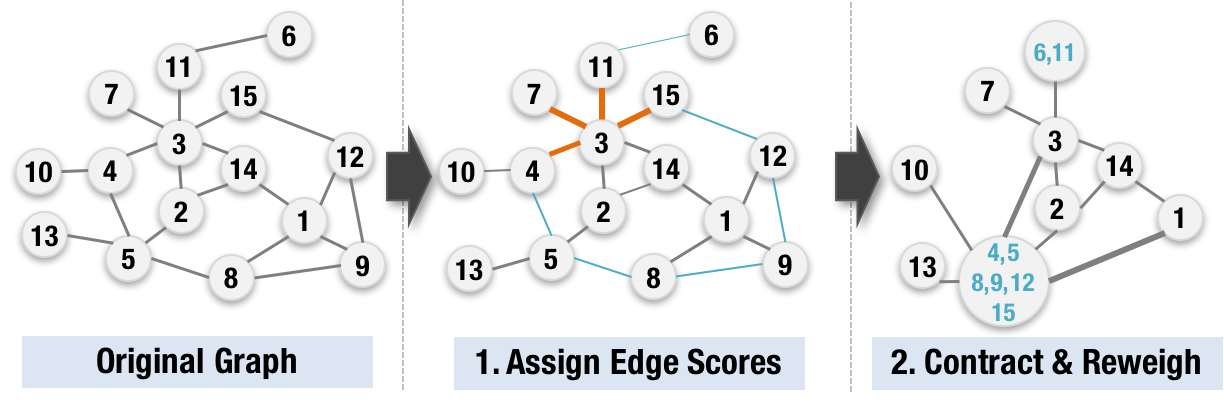}
\caption{\newnew{Overview of \coarsenet~\cite{purohit2014fast}. All the edges in the original graph are weighted equally. In step 1, edges with small width result in small changes in $\lambda_1$, while heavy edges result in big changes and are not good candidates for contraction. In step 2, the edge width depicts the new edge weight after obtaining the coarsened network.}}
\label{fig:coarsenet}
\end{figure}

The methods described above all minimize some version of the approximation or reconstruction error. \new{Other node-grouping approaches seek summaries that maintain specific properties of the original graph, a goal that resembles the target of graph sparsification methods~\cite{SpielmanS11,HublerKBG2008}. One example is diffusive properties related to the spectrum of the graph, and specifically its first eigenvalue $\lambda_1$~\cite{purohit2014fast}, which are crucial in diffusion and propagation processes like epidemiology and viral marketing. In this case, the summarization problem is formulated as a minimization of the change in the first eigenvalue between the adjacency matrices of the summary and the original graph. For efficiency, the method repeatedly merges pairs of \textit{adjacent} nodes, and uses a closed form to evaluate the change in $\lambda_1$, derived using matrix perturbation theory. 
Node pairs are merged in increasing order of change in $\lambda_1$---\newnew{the light edges with small ``edge scores'' in step 1 of Figure~\ref{fig:coarsenet} are good candidates for merging}---and the merging process stops when the user-specified number of nodes is achieved. 
At every step, edges are reweighted so that $\lambda_1$ is maintained \newnew{(step 2 in Figure~\ref{fig:coarsenet})}.  
The temporal extension of this approach is discussed in Section~\ref{sec:dynamic}. }

 
In the \new{visualization domain, \citet{DunneS13} introduce motif simplification to enhance network visualization.
Motif simplification replaces common links and common subgraphs, like stars and cliques, with compact glyphs to help visualize and simplify the complex relationships between entities and attributes. This approach uses exact pattern discovery algorithms to identify patterns and subgraphs, replacing these with glyphs to result in a less cluttered network display. We give an example in Section ~\ref{sec:app_viz}.} 

Beyond the end goal of summarization itself, node grouping can be applied to many graph-based tasks. 
CoSum~\cite{zhu2016unsupervised} involves summarization on $k$-partite heterogeneous graphs to improve record linkage between data sets, otherwise known as entity resolution. 
CoSum transforms an input $k$-type graph into another $k$-type summary graph composed of supernodes and superedges, 
using links between different types to improve the accuracy of entity resolution. 
The algorithm jointly condenses vertices into a supernode such that each supernode consists of nodes of the same type with high similarity,
and creates superedges that connect  supernodes according to the original links between their constituent nodes. The resultant summary achieves better performance in entity resolution than generic approaches, especially in data sets with missing values and one-to-many or many-to-many relations.

\subsubsection{\bf Edge-grouping methods} 
\label{subsec:edge-grouping}

\new{\textit{Unlike node-grouping methods that group nodes into supernodes, edge-grouping methods aggregate edges into \emph{compressor} or \emph{virtual nodes} to reduce the number of edges in a graph in either a lossless or lossy way.
Note that in this section, ``compression'' does not refer to bit-level optimization, as in the following section, but rather to the process of replacing a set of edges with a node.}}

\begin{wrapfigure}{L}{0.45\textwidth}
\centering
        \includegraphics[width=0.45\textwidth]{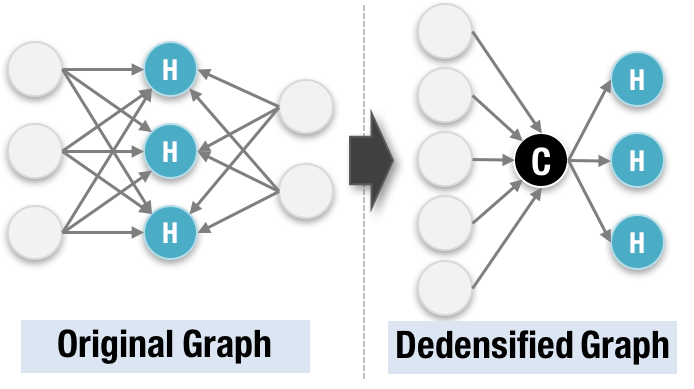}
        \caption{Example of Graph Dedensification~\cite{maccioni2016dedensification}: many edges are  removed after the addition of the compressor node \newnew{\textbf{C}, which connects to the high-degree nodes \textbf{H}.}}
        \label{fig:dedensification}
\end{wrapfigure}

Graph Dedensification~\cite{maccioni2016dedensification} is an edge-grouping method that compresses neighborhoods around high-degree nodes, accelerating query processing and enabling direct operations on the compressed graph. 
Following the assumption that high-degree nodes are surrounded by redundant information that can be synthesized and eliminated, \citeauthor{maccioni2016dedensification} introduce ``compressor nodes'', which represent common connections high-degree nodes. 
\new{To provide global guarantees and reduce the scope of compressor handling during query processing, 
dedensification only occurs when every node has at most one outgoing edge to a compressor node, and every high-degree node has incoming edges coming only from a compressor node.} 
These guarantees are then used to create query processing algorithms that enable direct 
pattern matching queries on the compressed graph. 

\new{Similar approaches include the ``connector'' motif in visualization-based summarization~\cite{DunneS13} discussed in Section~\ref{subsec:node-grouping} and 
Virtual Node Mining (VNM)~\cite{buehrer2008scalable}, which is used as a lossy compression scheme for the Web graph} to accommodate community-related queries and other random access algorithms on link servers. 
Like SUBDUE~\cite{cook:94:subdue} (Section~\ref{sec:compression_labeled}), VNM uses a frequent mining approach to extract meaningful connectivity formations by casting the outlinks/inlinks of each vertex as a transaction/itemset. Then, for each recurring pattern, it removes the links from its vertices and generates a new vertex (virtual node) in the graph, which is added as an outlink.
The process may be viewed exactly like graph dedensification (Figure~\ref{fig:dedensification}),
\new{although dedensification provides exact answers due to its losslessness and does not suffer from the space/time trade-off of graph indexing. }

\subsection{\new{Bit} compression-based methods}
\label{sec:compression}

\textit{Bit compression is a common technique in data mining. In graph summarization, the goal of these approaches is to minimize the number of bits needed to describe the input graph, where the summary consists of a model for the input graph and its unmodeled parts. 
The graph summary or model is significantly smaller than the original graph and often reveals various structural patterns, like bipartite subgraphs, that enhance understanding of the original graph structure. 
As mentioned in the previous section, some of these approaches primarily use compression and secondary grouping techniques.
However, some others aim solely to compress a given graph without necessarily creating a graph summary or finding comprehensible graph structures.} 

Here we focus mostly on the former approaches, which often formulate summarization as a model selection task.
These works employ the two-part Minimum Description Length (MDL) code, the goal of which is to minimize the description of the given graph $G$ and the model class $M$ in terms of bits:
\begin{equation}
min L(G,M) = L(M) + L(G | M),
\label{eq:mdl}
\end{equation}
which is given as the description length of the model, $L(M)$, and the description length of the graph given the model (i.e., the errors or unmodeled parts with respect to the model). 
For completeness, we also present some graph compression methods that can be adapted to summarization, although not originally designed for that purpose. 

\vspace{0.1cm}

\begin{wrapfigure}{l}{0.42\textwidth}
\centering
\includegraphics[width=0.42\textwidth]{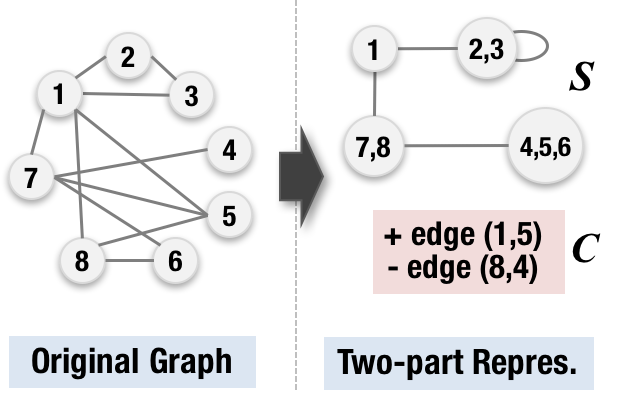}
\caption{\newnew{Two-part MDL representation~\cite{NavlakhaRS08}: graph summary $S$ and corrections $C$. Since $S$ does not capture the edge (1,5) properly, it is added in $C$. Similarly, the summary ``captures'' edge (8,4), which is missing in the original graph, so it is removed in $C$.}}
\label{fig:mdl}
\vspace{-0.1cm}
\end{wrapfigure}

Relying on this two-part MDL representation, \citet{NavlakhaRS08} 
introduce an approach to summarize graphs with bounded error. This representation, obtained by aggregating nodes in the summary generation, consists of a graph summary $S$ and a set of corrections $C$ (Figure~\ref{fig:mdl}). The summary is an aggregate graph in which each node corresponds to a set of nodes in $G$, and each edge represents the edges between all pairs of nodes in the two sets. The correction term specifies the list of edge-corrections that must be applied to the summary to exactly recreate $G$. The cost of a representation, $R$, is the sum of the storage costs of both $S$ and $C$: $cost(R) = |E_S| + |C|$, where $E_S$ is the set of superedges in $S$. \new{The MDL-based graph summary is found by aggregating groups of nodes (thereby falling also into the grouping-based summarization category) as long as they decrease the MDL cost of the graph. To this end, a simple but costly greedy heuristic iteratively combines node pairs that give the maximum cost reduction into supernodes. To reduce the complexity to cubic on the average degree of the graph, a randomized algorithm randomly picks a node and merges it with the best node in its 2-hop neighborhood. This formulation also supports lossy compression with bounded reconstruction error in order to achieve even higher space savings.} This summarization approach gives up to two times more compact summaries than graph compression~\cite{boldi04webgraph} and clustering~\cite{graclus} methods. 

\new{Similar to \citet{NavlakhaRS08}, \citet{ahnert2013power} introduces a biological application for the discovery of dominant relationship patterns in transcription networks, such as the networks of \textit{S. serevisiae} and \textit{E. coli}. In biology, the terms ``power graph'' and ``power nodes/edges'' are used to refer to what we call supergraphs and supernodes/edges. In this application, 
most supernodes are shown to have functional meaning, and the superedges signify large-scale functional relationships between different subsystems of transcription networks.}


\begin{figure}[t!]
\centering
        \includegraphics[width=1.0\textwidth]{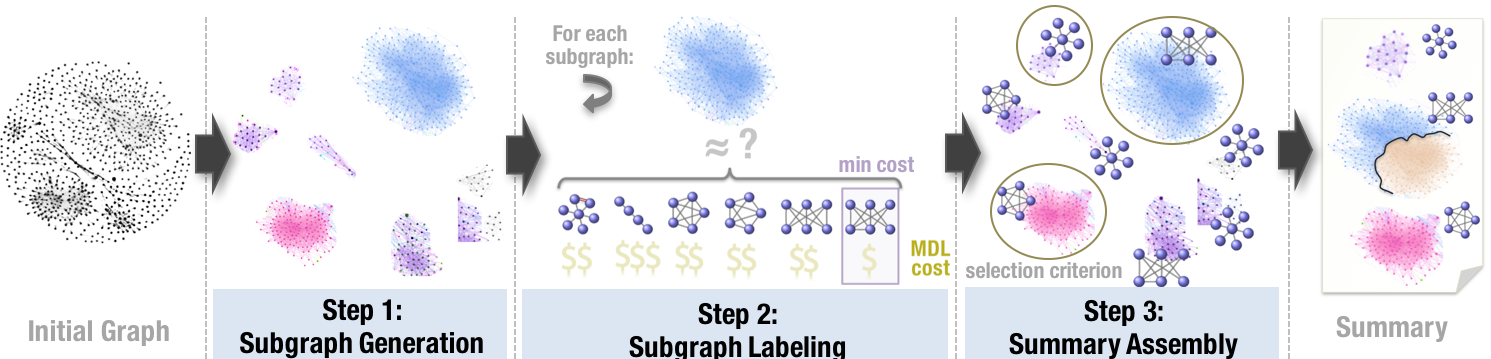}
        \caption{\vog~\cite{koutra:14:vog}: \newnew{Overview of vocabulary-based graph summarization.}}
        \label{fig:vog}
\end{figure}

Addressing an information-theoretic optimization problem also based on MDL, \vog~\cite{koutra:14:vog}, or \emph{vocabulary-based summarization of graphs}, succinctly describes large-scale graphs with a few possibly overlapping, easily understood structures  encoded in the model $M$. 
The graph summary is given in terms of a predefined ``vocabulary" of structures that goes beyond the simple rectangles that most summarization and clustering methods find, identifying cliques and near-cliques, stars, chains, and (near-) bipartite cores. \vog is modular (Figure~\ref{fig:vog}): (i)~it first performs graph clustering by adapting the node reordering method \slashburn~\cite{6807798} to extract ego-networks and other disconnected components; (ii)~it labels the extracted subgraphs with the appropriate structures in the assumed vocabulary (i.e., cliques and near-cliques, stars, chains, and full or near-bipartite cores) using MDL as a model selection criterion; and (iii)~finally it creates a summary by employing heuristics that choose only the subgraphs that minimize the total encoding cost of the graph, $L(G,M)$, as it is defined in Equation~\eqref{eq:mdl}. \new{Some of the exact structures in the vocabulary are part of the motif simplification scheme~\cite{DunneS13} (Section~\ref{subsec:node-grouping}), but \vog is distinct in that it allows for near-structures that appear often in real-world graphs and uses MDL for summarization.
Likewise, \vog and~\citeauthor{NavlakhaRS08}'s MDL representation are similar in that they use MDL for summarization, but the latter is confined to summarizing a graph in terms
of non-overlapping cliques and bipartite cores, while \vog supports a more diverse set of structures or vocabulary. Moreover, it is possible to expand the vocabulary to address the needs of specific applications or domains.} Extensions of \vog~\cite{liu2015empirical} have been applied to empirically evaluate the summarization power of various graph clustering methods, such as METIS~\cite{Karypis99@METIS}. Similar to \vog, ~\cite{miettinen:11:mdl4bmf,miettinen:14:mdl4bmf} discuss MDL for Boolean matrix factorization, which can be viewed as a summary in terms of possibly overlapping full cliques in directed graphs. 

\vspace{0.1cm}
\noindent \textbf{Connections to graph compression.}
Graph summarization and compression are related. \new{Graph summarization methods leverage compression in order to find a smaller representation of the input graph, \textit{while discovering structural patterns}.  In these cases, although compression is the means, finding the absolutely smallest representation of the graph is \textit{not} the end goal. The patterns that are being unearthed during the process may lead to suboptimal compression. On the other hand, in graph compression works, the goal is to compress the input graph as much as possible in order to minimize storage space, irrespective of patterns.} 

\new{Since compression and summarization are distinct fields, we only give a few fundamental methods in the former, including: the so-called ``Eulerian data structure'' to handle neighbor queries in social networks~\cite{MaserratP10} and extensions of this work to community-preserving compression~\cite{maserrat2012community}; 
node reordering techniques, such as zip block encoding in \gbase~\cite{kang2011gbase}, bipartite minimum logarithmic arrangement~\cite{dhulipala2016compressing} for inverted indices, and techniques for real graphs with power-law degree distributions~\cite{6807798}; edge reordering techniques~\cite{GoonetillekeKSL17}; compression of web graphs using lexicographic localities~\cite{boldi04webgraph}; extensions to social networks~\cite{grabowski2014tight,ChierichettiKLMPR09}; breadth first search-based approaches~\cite{ApostolicoD09}; lossy edge encoding per triangle~\cite{feng:13:badmdl}; weighted graph compression to maintain edge weights up to a certain number of hops~\cite{ToivonenZHH11}; provably optimal compression of Erd\"{o}s-R\'{e}nyi random graphs using structural entropy (SZIP~\cite{choi2012compression}); and minimal probabilistic tile cover mining~\cite{liu2016summarizing} that has applications to binary matrices and bipartite graphs.}

\subsection{Simplification-based methods}
\label{sec:simplification} 

\textit{Simplification-based summarization methods streamline the original graph by removing less ``important" nodes or edges, resulting in a sparsified graph. As opposed to supergraphs, here the summary graph consists of a subset of the original nodes and/or edges.
In addition to simplification-based summarization methods, some existing graph algorithms have the potential for simplification-based summarization, such as sparsification, sampling, and sketching.}

A representative work on \textit{node} simplification-based summarization techniques is OntoVis~\cite{shen2006visual}, a visual analytical tool that relies on node filtering for the purpose of understanding large, heterogeneous social networks in which nodes and links respectively represent different concepts and relations. 
OntoVis uses information in the ontology that relates nodes and edges, such as the degree of nodes of specific type, to semantically prune the network.
OntoVis supports semantic abstraction, structural abstraction, and importance filtering. 
In semantic abstraction, the user constructs a derived graph from the original graph by including only nodes whose types are selected from the original ontology graph.
For example, in a terrorism network, selection of the node type ``terrorist organization'' results in a semantic abstraction of different terrorist organizations. 
Structural abstraction simplifies the graph while preserving the essential structure of the entire network, for example by removing one-degree nodes and duplicate paths. 
Importance filtering  makes use of statistical measures like node degree for evaluating connectivity and relevance between node types. 

\begin{figure}[t]
\centering
\includegraphics[width=0.95\textwidth]{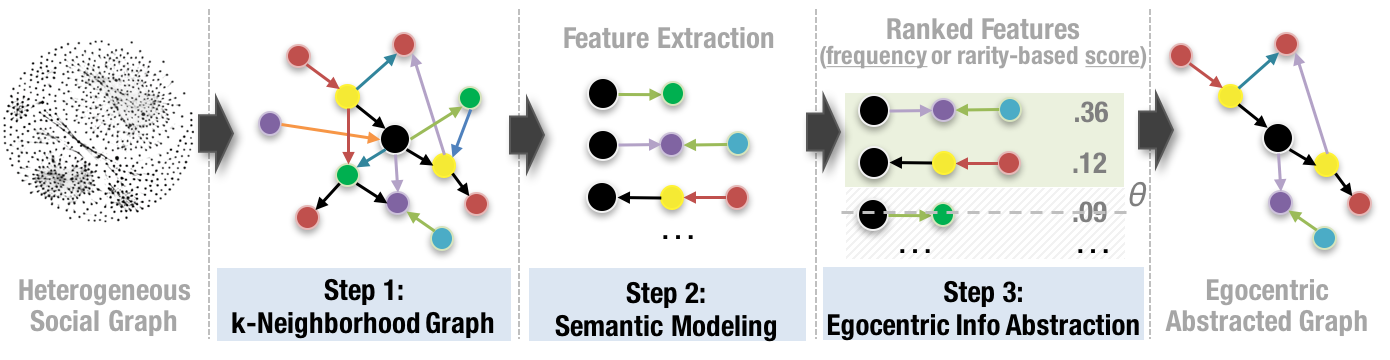}
\caption{Overview of egocentric abstraction~\cite{li2009egocentric}. \newnew{The features  are ranked by frequency (score) in step 2. Depending on the policy, only the frequent or rare features are used. In the example, the abstraction graph is based on frequent features above a threshold $\theta$.} 
}
\label{fig:egocentric}
\vspace{-0.4cm}
\end{figure}

Targeting the same type of graph as OntoVis, \citet{li2009egocentric} propose a four-step unsupervised algorithm for egocentric information abstraction of heterogeneous social networks using edge, instead of node, filtering (Figure~\ref{fig:egocentric}). First, during the semantic modeling step, features (or else linear combinations of relations, or path-based patterns) are automatically selected and extracted according to the surrounding network substructure, ($k$-hop neighborhoods). Second, the statistical dependency is measured between the features per ego node. 
Third, during the egocentric information abstraction step, irrelevant information is removed by applying distilling criteria, such as keeping the most frequent or rare features. 
Finally, in the fourth step, an egocentric abstracted graph is constructed incrementally on the remaining features, allowing the user to visualize the smaller resulting graph.  

\vspace{0.1cm}
\noindent {\bf Connections to graph sampling, sparsification, and sketches.} 
\new{A complementary approach toward ``compressing'' a graph involves sampling nodes or edges from it~\cite{HublerKBG2008,BatsonSST13}. Note, though, that sampling focuses more on obtaining sparse subgraphs that can be used to approximate properties of the original graph (degree distribution, size distribution of connected components, diameter, or community structure~\cite{MaiyaB10}) and less on identifying patterns that collectively summarize the input graph to enhance user understanding.} 

Various sampling techniques have been studied ~\cite{MathioudakisBCGU11,AhmedNK13}, and a comprehensive tutorial on graph sampling was presented at KDD~\cite{networksampling}. Sampling techniques include sampling nodes according to their in- or out-degree, PageRank, or substructures, such as spanning trees; \new{as well as sampling edges uniformly, or according to their weights or their effective resistance~\cite{SpielmanS11} to maintain the graph spectrum up to some multiplicative error or to maintain node reachability (transitive reduction~\cite{AhoGU72}).} Although sampling has the potential to allow better visualization~\cite{RafieiC05} and approximate specific queries with theoretical guarantees, \new{it cannot detect graph structures, often operates on individual nodes/edges instead of collective patterns,}  and may need additional processing in order to make sense of the sample.  
Related to the goal of maintaining specific graph properties is the $k$-spanner~\cite{PelegS89}, which is the sparsest subgraph in which the distance between pairs of nodes is at most $k$ times the distance in the initial graph. A common category of the problem is the tree $k$-spanner, which approximates the original graph with a tree that satisfies the distance property. Finding a $k$-spanner is NP-hard except for the case of $k = 2$, which can be solved in $O(|\mathcal{E}|+|\mathcal{V}|)$ time.

Graph sketches~\cite{AhnGM12,Liberty13,ghashami2016efficient}, or data synopses obtained by applying linear projections, are also relevant. Graph sketching can be viewed as linear dimensionality reduction, where the linearity of sketches makes them applicable to the analysis of streaming graphs with node and edge additions and deletions and distributed settings, such as MapReduce~\cite{Dean04MapReduce}.

\subsection{Influence-based methods}
\label{subsec:infl_static}
\textit{\newnew{Influence-based methods seek to find a compact, high-level description 
of the influence dynamics 
in large-scale graphs, in order to understand the patterns of influence propagation at a global level}. Usually such methods formulate graph summarization as an optimization process in which some quantity related to information influence is maintained.  \new{These summarization methods are scarce and have been mostly applied on social graphs, where important influence-related questions arise}.}

Community-level Social Influence or CSI~\cite{Mehmood2013} is a representative work that focuses on summarizing social networks via information propagation and social influence analysis. 
Like some other graph summarization techniques, CSI relies on existing clustering approaches: it detects a set of communities using \metis~\cite{Karypis99@METIS} and then finds their reciprocal influence by extending the popular Independent Cascade model~\cite{Kempe03Maximizing} to communities instead of individual nodes. 
To balance between data fit and model complexity, CSI uses MDL and Bayesian Information Criterion (BIC) approaches to select the number of communities for the graph model. 
Unlike influence propagation approaches that find representative cascades for information diffusion, CSI leads to a compact representation of the input network where the nodes correspond to communities and the directed edges represent influence relationships. 
Note that the output of CSI is different from grouping-based summarization techniques in which the superedges simply represent aggregate connections between the adjacent supernodes.
\new{SPINE, an alternative to CSI~\cite{MathioudakisBCGU11}, sparsifies social networks to only keep the edges that ``explain" the information propagation---those that maximize the likelihood of the observed data. This problem is shown to be NP-hard to approximate within any multiplicative factor. 
Inspired by the idea of decomposing sparsification into a number of subproblems equal to the nodes in the network, SPINE is a greedy algorithm that achieves efficiency with practically little compromise in quality.} Unlike CSI, it simply eliminates original edges and does not group nodes into communities or supernodes.

\subsection{\new{Other Types of Graph Summaries}}

\begin{figure}[t!]
          \centering
          \includegraphics[width=0.95\textwidth]{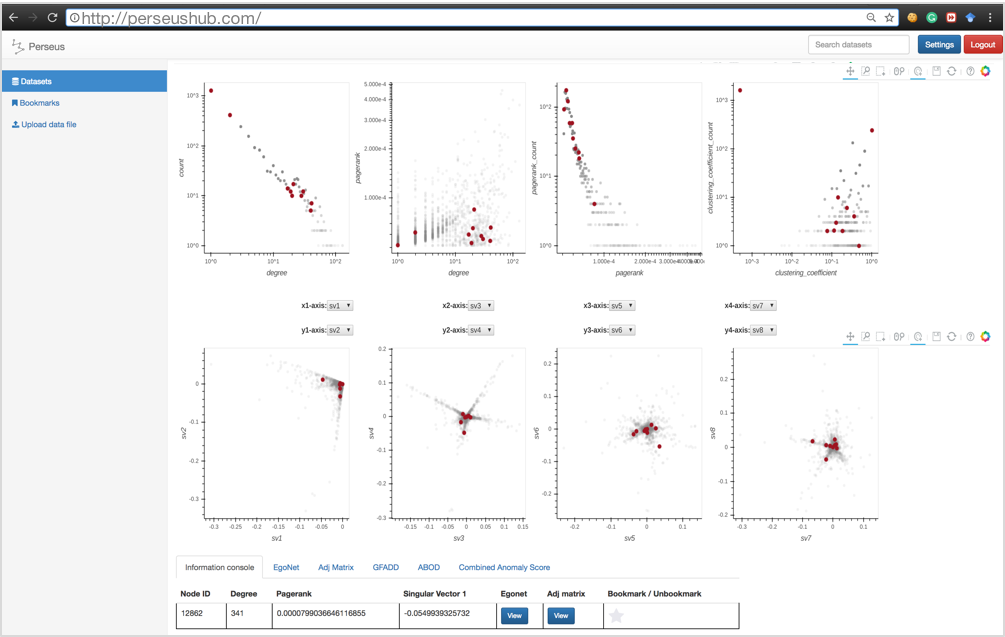}
          \caption{The front-end of \textsc{Perseus-Hub}, with linked plots for graph properties. The annotated red points correspond to 
          anomalies found during  offline pre-processing.}
          \label{fig:perseus_frontend}
          \vspace{-0.4cm}
\end{figure}
\new{Although not our main focus, we briefly present methods that represent a network (i) visually with a small set of anomalous patterns, distribution plots of graph properties, or carefully selected nodes, or (ii) with latent representations.} 

\noindent \new{\textbf{Visualization-based systems.}} Various graph visualization platforms for pattern identification exist. 
For example, Apolo~\cite{ChauKHF11} routes attention by visualizing the neighborhoods of a few user-selected seed nodes, which can be interactively explored. 
A follow-up anomaly detection system, OPAvion~\cite{akoglu2012opavion},  
mines graph features using the Hadoop-based graph mining framework Pegasus ~\cite{PegasusICDM2009}, spots anomalies by employing OddBall~\cite{AkogluMF10} for mining distributions of egonet-related features (e.g., number of nodes vs.\ edges), and interactively visualizes the anomalous nodes via Apolo. 
Finally, the large-scale system \perseus~\cite{KoutraJNF15,JinLSZWK17}  enables comprehensive graph analysis by supporting the coupled summarization of graph properties (computed on Hadoop or Spark) and structures, guiding attention to outliers, and allowing the user to interactively explore normal and anomalous node behaviors in distribution plots and ego-network representations (Figure~\ref{fig:perseus_frontend}).
Other visualization-based methods include scaled density plots to visualize scatter plots~\cite{Shneiderman08}, random and density sampling~\cite{BertiniS04_2} for datasets with thousands of points,  
and rescaled visualization of spy, distribution, and correlation plots of massive graphs~\cite{KangLKF14}.

\new{Visualization-based graph summarization is also related to visual graph analytics in that summaries of graphs can support interactive visualization. However, the traditional focuses of visual graph analytics, such as the layout of the data displayed and new visualization or user interaction techniques, differ from the typical goals of graph summarization. 
Widely used visualization tools, such as Gephi~\cite{gephi}, Cytoscape~\cite{shannon2003cytoscape}, and the Javascript D3 library~\cite{bostock2011d3}, support interactive  exploration of networks and operations such as spatializing, filtering, and clustering. Although these platforms work well on small and medium-sized graphs, they cannot render large-scale networks with many thousands or millions of nodes, or else they are compromised by high latency. These tools can benefit from graph summarization methods that result in smaller network representations or patterns thereof, which can be displayed more easily. }

\vspace{0.05cm}
\noindent \newnew{\textbf{Domain-specific Summaries.} Beyond visualization, \citet{JinK17eagle}
propose an \textit{optimization} problem for summarizing a graph in terms of representative, domain-specific graph properties. The summaries are required to be concise, diverse, domain-specific, interpretable, and fast to compute. This is the first work to target domain-specific summarization by automatically leveraging the knowledge encoded in multiple networks from a specific domain, like social science or  neuroscience.
Although it is related to visualization-based systems that support coupled summarization of graph properties (e.g., \perseus~\cite{KoutraJNF15,JinLSZWK17}, described above), this method automates the selection of the graph properties to be included in the graph `summary' based on the domain from which the data comes.} 

\vspace{0.05cm}
\noindent \new{\textbf{Latent Representations.}} \newnew{A variety of methods 
obtain low-dimensional representations of a network in a latent space.} For instance, matrix factorization methods, such as SVD, CUR~\cite{V-:Drineas2006}, and CMD~\cite{Sun07@SDM}, all lead to low-rank approximations of an adjacency matrix, which \newnew{can be viewed as sparsified approximate} ``summaries''  of the original graph. 
Recent interest in deep learning has lead to novel \textit{node} representation learning techniques (e.g.,~\cite{deepwalk,node2vec,WangCZ16,TangQWZYM15}), but these methods present nodes as low-dimensional vectors instead of finding a compact graphical representation of the whole network, which is the goal of summarization.

\section{Static Graph Summarization: Labeled Networks}
\label{sec:attribute}
So far we have reviewed summarization methods that use the structural properties of static graphs without additional information like node and edge attributes. However, many real graphs are annotated, labeled, or attributed. 
For example, in a social network, a typical node representing a user is associated with information about age, gender, and location; transportation graphs may have information about the capacity of streets (edges) and the maximum speed per street; forums like Quora, which can be interpreted as networks of questions and answers, have comments, upvotes, and downvotes. 
A general definition of graph summarization for static, labeled graphs is given as follows: 

\begin{problem} \textbf{Summarization of Static, Labeled Graphs}. \hfill \\ 
\textbf{Given} a static graph $\G$  
 and side information, such as node attributes, \\
{\bf Find} a labeled summary graph or a set of labeled structures 
to {\bf concisely describe} the given graph.
\end{problem}

Overall, the main challenge in summarizing labeled graphs is the efficient combination of two different types of data: structural connections and attributes.  
Currently, most existing works focus on node attributes alone, 
although other types of side information are certainly of interest in summarization. For instance, joint summarization of multimodal data---including graphs, text, images, and streaming data---has various applications.
However, \newnew{due to the challenges of multimodal analysis}, these methods are underexplored in the literature.

The second block of Table~\ref{tab:static_qualitative} provides qualitative comparisons and explicit characterizations of static graph summarization methods for labeled graphs, which we review next by classifying them based on their core technical methodology. The overview of this section is included in Figure~\ref{fig:taxonomy}.

\subsection{Grouping-based methods}
\label{sec:grouping_labeled}

\textit{Grouping-based methods aggregate nodes into supernodes connected by superedges based on both structural properties and node attributes. Grouped nodes are usually structurally close in the graph and share similar attribute values.} 

\new{As discussed with plain graphs, here attributed clustering or community detection~\cite{zhou2009graph,yang2013community,Xu:2012:MAA:2213836.2213894} methods do not perform summarization, but could be leveraged by summarization approaches to obtain compact representations of graphs with attributes. One fundamental difference between summarization and clustering is that the former finds coherent sets of nodes with similar connectivity patterns to the rest of the graph, while clustering results in coherent, densely-connected groups of nodes.} 

\begin{wrapfigure}{r}{0.32\textwidth}
\vspace{-0.55cm}
    \centering
    \includegraphics[width=0.28\textwidth]{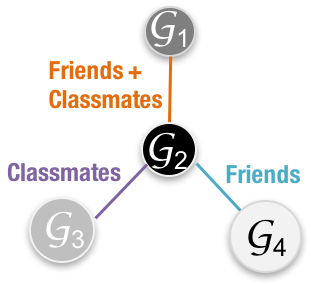}
    \caption{\newnew{SNAP summary~\cite{tian2008efficient} of a student graph: each student in $\mathcal{G}_1$ has at least one friend and one classmate in $\mathcal{G}_2$. The node size reflects the number of people per group $\mathcal{G}_i$.}    
    \label{fig:snap}}
    \vspace{-0.2cm}
\end{wrapfigure}
Optimizing specifically for Web graphs, the S-Node representation~\cite{raghavan2003representing} is a novel two-level lossless graph compression scheme. Here a Web graph is a set of small directed graphs consisting of supernodes and superedges, which are pointers to lower-level graphs that encode the interconnections within a small subset of Web pages. 
S-Node exploits empirically observed properties of Web graphs like domain locality and page similarity, \new{some of which can be viewed as node labels and others as additional textual information,} to guide the grouping of pages into supernodes. 
Using a compression technique called reference encoding for the lower level directed graphs, S-Node achieves high space efficiency and naturally isolates portions of Web graphs relevant to particular queries. 
This representation is the first Web graph representation scheme to combine compression with support for both complex queries and local graph navigation.

Mostly studied in the database community, grouping-based attributed graph summarization methods tend to rely on operations related to GROUP BY. SNAP and $k$-SNAP are two popular database-style 
approaches~\cite{tian2008efficient}. 
\new{SNAP relies on (A,R)-compatibility (attribute- and relationship-compatibility), which guarantees that nodes in all groups are homogeneous in terms of attributes, and are also adjacent to nodes in the same 
groups for \textit{all} types of relationships.
For example, in Figure~\ref{fig:snap}, each student in $\mathcal{G}_1$ has at least one friend and classmate in $\mathcal{G}_2$.  
 SNAP begins by creating groups of nodes that share the same attributes, and then iteratively splits these groups until the grouping is ``compatible'' with the relationships, eventually producing the maximum (A,R)-compatible grouping.}
The nodes of the summary graph given by SNAP correspond to the groups, and the edges are the group relationships. 
$k$-SNAP further allows users to control the summary resolution, providing ``drill-down'' and ``roll-up'' capabilities to navigate through summaries of different resolutions. 

In order to facilitate interactive summarization, CANAL~\cite{ZhangTP10} automates $k$-SNAP by categorizing numerical attribute values, exploiting domain knowledge about the node labels and graph structure. 
To point users to the potentially most useful summaries, CANAL incorporates three  ``interestingness'' criteria: 
(i)~\textit{Diversity}, the number of strong relationships connecting groups with different attribute values; 
(ii)~\textit{Coverage}, the fraction of nodes in the original graph that are present in strong group relationships; (iii)~\textit{Conciseness}, the sum of the number of groups and strong group relationships, where a lower sum is preferred. Overall, interestingness is given as $\frac{Diversity(S) \times Coverage(S)}{Conciseness(S)}$, where $S$ is the summary graph.

\citet{hassanlou2013probabilistic} introduce another database-centered graph summarization approach similar to SNAP, where each node group consists of nodes that have the \textit{same} attribute values using the GROUP BY operation. 
Unlike SNAP, though, this approach applies to probabilistic graphs, or graphs with edges that have probabilities of existence associated with them. 
\citet{shoaran2013zero} extend this by aiming to protect the privacy of data in the labeled summaries generated by the aforementioned probabilistic technique. 
Finally, \citet{gehrke2011zkp} propose a privacy framework that extends Zero-Knowledge Privacy, improving upon differential privacy by only considering a random sampling of data with added noise for the summarization. 

In the database community, \citet{fan2012query} propose a ``blueprint'' for lossless queries on compressed attributed graphs.
To achieve this, query-specific functions are introduced for compressing the graph, rewriting the query accordingly, and interpreting the result of the rewritten query on the compressed graph. 
For example, this blueprint can be implemented for queries of reachability (i.e., can node $A$ be reached from node $B$?) and pattern matching (i.e., is there a subgraph that best satisfies a function provided by the user on path length between nodes in the subgraph?). \new{The key idea is to group nodes that belong to the same equivalence class; intuitively, nodes that are similar in structure and labels are equivalent.}
This differs from other database-style operations that first group nodes by labels and later analyze the structure. 
To handle dynamic changes in web, social and other networks, the authors also introduce unbounded algorithms that evaluate incremental graph structure changes and propagate the changes to the compressed graph representation.
\citet{RenW15} propose a method similar to \citeauthor{fan2012query} specifically for subgraph isomorphism queries, where  groupings are based not only on equivalent nodes but also on edge-specific relationships that optimize the vertex matching order. 


\newnew{In the case of schema-less databases---in particular, for knowledge graphs connecting entities and concepts---\citet{song2016knowledgegraphs} propose a lossy graph summarization framework as a collection of \emph{$d$-summaries}, which intuitively are supergraphs that group similar entities (i.e., with the same attribute or label) within $d$ hops of each other. Specifically, the entities within a $d$-summary observe what is called \emph{$d$-similarity}, which preserves directed paths up to length $d$. 
Unlike frequent subgraph mining---a building block for various graph algorithms, including summarization---which is  NP-hard, computing $d$-summaries is tractable. 
To evaluate $d$-summaries, the authors introduce approximations of an NP-hard ``bi-criteria'' function that quantifies \emph{informativeness} and \emph{diversity}.
The former measure favors large summaries with high coverage of the original graph; the latter penalizes redundancy for entities appearing in many $d$-summaries.
 Both summarizing and querying knowledge graphs with  $d$-summaries are efficient, and can maintain up to 99\% accuracy for subgraph queries in real and synthetic graphs.} 

\new{Beyond attribute- and relationship-coherent summaries, there exists work on creating summaries from frequently occurring subgraphs in heterogeneous labeled graphs. A representative work is dependence graph summarization, where the vertices are labeled with program  operations and the edges represent dependency relationships between them~\cite{chen2009mining}. 
The algorithm first generates partitions created by sampling nodes of the same label, resulting in multiple groups with consistent labels.
The partitioning/summarization is followed by frequent subgraph mining and verification (removal of false positives). These steps are performed in multiple iterations to find a lower bound on the false negative rate of frequent subgraph detection.}

\subsection{\new{Bit} compression-based methods}
\label{sec:compression_labeled}

\textit{Most compression-based summarization methods leverage  MDL  to guide the grouping of nodes or the discovery of frequent structures to be replaced with virtual nodes in the summary. Here, the employed compression and/or aggregation techniques consider both the graph structure and node/edge attributes.}

The first and most famous frequent-subgraph-based summarization scheme, SUBDUE~\cite{cook:94:subdue}, employs a two-part MDL representation (described in Section~\ref{sec:compression}). Beyond the network structure, the MDL encoding accounts for node and edge labels. Greedy beam search is used to iteratively replace the most frequent subgraph in a labeled graph, which minimizes the MDL cost, with a meta-node.
Multiple passes of SUBDUE eventually produce a hierarchical description of the structural regularities in the graph. The resulting representation can be used to either identify anomalous structures (instances that do not compress well) or the most common substructures (substructures that have very low compression cost). Since the introduction of SUBDUE, many methods have been proposed to alleviate the complexity issues of frequent pattern mining on graphs, or to extend its application in different settings: \citet{Maruhashi:2011} propose \emph{MultiAspectForensics}, a tool to detect and visualize graph patterns; 
\citet{Thomas:2010} introduce MARGIN, an algorithm that reduces the search space of frequent subgraphs by only mining the maximal frequent subgraphs of a graph database; and 
\citet{Wackersreuther:2010} propose a frequent subgraph mining algorithm to operate on dynamic graphs. 
\new{Similar to SUBDUE, a grammar-based compression scheme~\cite{maneth2016compressing} recursively replaces frequent ``substructures'' in directed edge-labeled hypergraphs, like RDF graphs. Rather than frequent subgraphs, these substructures are digrams, or pairs of connected hyperedges: for example, the digram ``ab'' consists of the edge labels ``a'' and ``b''. 
The process of recursive replacement of digrams stops when no digram occurs more than once. Unlike most compression-based works that use MDL, this approach leverages variable-length $\delta$-codes~\cite{Elias06} for the connectivity  and edge labels.}

\new{A simpler information-theoretic approach that does not use frequent subgraph mining directly minimizes the two-part MDL representation of an input network~\cite{wu2014graph}.}
The model cost consists of the number of bits to describe three parts: the number of node and attribute groups; the nodes in each group; and the links among groups. The data cost includes the description cost of the links inside each group and the attributes.
The greedy summary-generating algorithm employs the MDL cost function to determine whether a certain node grouping is beneficial to the summary as a whole (i.e., it reduces the total encoding cost of the graph). A faster version of the greedy algorithm initializes the summaries using label propagation instead of random initialization.

\new{Beyond its standalone utility, MDL can be easily combined with other techniques, such as locality-sensitive hashing (LSH)~\cite{AndoniI08}, to help with in-memory processing and summary generation.} LSH is a popular technique for efficient similarity search (here, nodes in the graph setting). \new{In the context of summarization, it can operate on the structure and labels of each node in order to efficiently find similar nodes that can be aggregated into a ``coherent'' group.} \citet{khan2014set} propose to LSH-based graph summarization by iteratively computing minhash functions on node neighborhoods, combining these minhash functions into groups, computing hash codes on the groups, and then aggregating the nodes that have the same hash codes.
To handle the labels in the graph,  adjacency and attribute lists are concatenated together before hashing. Supernodes are used to combine nodes, and, unlike other works, virtual nodes are used to combine edges between groups of nodes. Here, MDL is used to measure the relative increases in compression efficiency achieved by grouping nodes to supernodes and edges to superedges.

We further note that MDL is used frequently for data that, while not explicitly modeled as a graph, can be implicitly viewed as such: R-KRIMP and RDB-KRIMP~\cite{koopman:08:rkrimp,koopman:09:rdbkrimp} summarize multi-relational data, which can be viewed as attributed graphs. The former, R-KRIMP, finds characteristic patterns in single data tables, then finds a small set of multi-relational characteristic item sets within this reduced search space. The latter extends R-KRIMP by finding more expressive patterns.

\subsection{Influence-based methods}

\textit{Influence-based summarization methods for labeled graphs are currently scarce. The representative method in this category leverages both structural and node attribute similarities to summarize the influence or diffusion process in a large-scale network.}

The sole work in this category, VEGAS~\cite{shi2015vegas}, summarizes influence propagation in citation networks via a matrix decomposition-based algorithm. 
The summarization problem aims to find the community membership matrix $H$ of the nodes (papers in the citation network) such that $\min_{H\geq0}||M^G-HH^T||_F^2$, where $M^G=\frac{AA^T+A^TA}{2}$ is the node similarity matrix and $A$ is the adjacency matrix. In the case of labeled networks, $M^G$ is replaced with the generalized similarity matrix $M^D=\frac{(A\bigodot A^D)(A\bigodot A^D)^T+(A\bigodot A^D)^T(A\bigodot A^D)}{2}$ to incorporate side information.
Here, \new{$\bigodot$ indicates the Hadamard or element-wise product of matrices,} and $A^D$, which may be specified by the user, encodes pairwise attribute similarity between nodes.
In more detail,  
first the maximal influence graph $G$ is computed from the input influence graph $I$ by a rooted graph search that follows the standard BFS/DFS implementation 
from source node $f$. Then the matrices $M^G$, $A^D$, and $M^D$ are generated.
Finally non-negative matrix factorization is used to solve the above optimization, yielding the community membership matrix $H$. Nodes are assigned to clusters according to the maximum value in each row of $H$. Summaries are generated after link pruning, which is performed to select the $l$ best flows (links) for the final summary, dropping all other links.

\section{Dynamic Graph Summarization: Plain Networks}
\label{sec:dynamic}
Analyzing large and complex data is challenging by itself, so adding the dimension of time makes the analysis even more challenging and time-consuming. Despite this, most networks realistically do change over time: for example, communication patterns with others via phone or social networks; the connection between servers in a network; the flow of information, news and rumors; the distance between connected vehicles; the information transmitted between devices in a smart home environment. 

For this reason, the temporal graph mining literature is rich, mostly focusing on: laws and patterns of graph evolution in~\cite{DBLP:conf/icml/LeskovecF07,FerlezFLMG08,DBLP:conf/kdd/LeskovecBKT08,DBLP:conf/kdd/LeskovecKF05,DBLP:journals/datamine/SunTHFE08} and a comprehensive survey  by \citeauthor{CharuEvolSurvey14} in \citeyear{CharuEvolSurvey14}; 
anomaly and change detection in streaming graphs \cite{aggarwal2005online} or time-evolving networks~\cite{FerlezFLMG08,KoutraVF13,KoutraSVGF15}
; 
discovery of dense temporal cliques and bipartite cores using PARAFAC tensor decomposition and MDL (\cite{sun2007graphscope,araujo2014com2,koutra2012tensorsplat}); 
mining of cross-graph quasi-cliques~\cite{Pei@KDD05}; 
clustering using incremental static clustering~\cite{xu2011tracking} or a probabilistic approach based on mixed-membership blockmodels~\cite{FuLX09}; 
sampling of streaming graphs~\cite{AhmedNK13} and 
role discovery~\cite{conf/kdd/HendersonGETBAKFL12,RossiGNH12}. 

\enlargethispage{\baselineskip}
In this section, we focus on methods that summarize time-evolving networks (third block in Table~\ref{tab:static_qualitative}). Summarization techniques for time-evolving networks have not been studied to the same extent as those for static networks, possibly because of the new challenges introduced by the dimension of time. The methods are sensitive to the choice of time granularity, which is often chosen arbitrarily: depending on the application, granularity can be set to minutes, hours, days, weeks, months, years, or some other unit that makes sense in a given setting. The continuous and sometimes irregular change of real-world graphs also complicates evolution tracking, defining online ``interestingness'' measures, and visualization. 
The dynamic graph summarization problem may be defined as:

\begin{problem} \textbf{Summarization of Dynamic, Plain Graphs}. \hfill \\
\textbf{Given} a dynamic graph, which is observed  as a set of streaming edges, or a sequence of adjacency matrices $\A_1, \A_2, \ldots, \A_T$ corresponding to the static graphs $\G_1, \G_2, \ldots, \G_T$ \\
{\bf Find} a temporal summary graph or  a set of possibly overlapping temporal structures to {\bf concisely describe} the given dynamic graph. 
\end{problem}

The summary is a time-evolving supergraph with supernodes and superedges, or else a sequence of sparsified graphs  with fewer nodes/edges than the input dynamic graph.

The simplest approach treats a time-evolving graph as a series of static graph snapshots, which allows the application of static graph summarization techniques on each snapshot. However, the effectiveness of this approach  depends  heavily on user-specified aggregation operations and the time granularity~\cite{SoundarajanTKECGR16}, and there is no globally established method for picking the ``right'' time unit. 
With small time granularity, the amount of data increases significantly. 
With large time granularity, interesting dynamics may be missed. 
Moreover, real-world processes can be unpredictable or bursty. Adjusting the time unit of analysis may be the key to understanding and capturing the important dynamics. 

\begin{wrapfigure}{R}{0.65\textwidth}
\vspace{-0.3cm}
\centering
\includegraphics[width=0.65\textwidth]{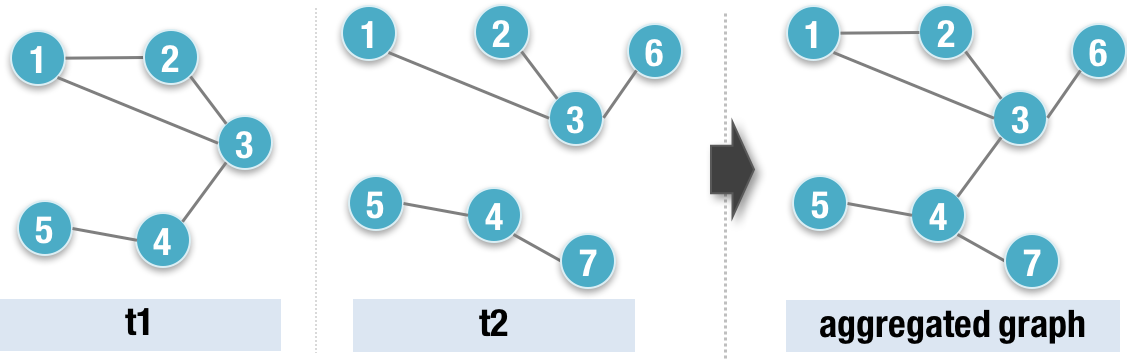}
\caption{Aggregated graph example (time t1+t2).}
\label{fig:aggregated}
\vspace{-0.2cm}
\end{wrapfigure}

An alternative  is to create an aggregate graph that summarizes the input dynamic network based on the recency and frequency of interactions (Figure ~\ref{fig:aggregated}). 
This has been called an ``approximation graph'' \cite{CortesPV02,HillABV06,SharanN08}. Specifically, the interactions between nodes in an approximation graph are aggregated over time and weighted by applying kernel smoothing (e.g. exponential, inverse linear, linear, uniform), where more recent edges are weighted higher than old edges.
Edges with weight below a specified threshold can also be pruned to simplify the graph approximation.
The approximation graph has been shown to be useful for telecommunications fraud detection~\cite{CortesPV02}, anomaly detection and prediction of user behavior in web logs and email networks~\cite{HillABV06}, and attribute classification via relational classifier models~\cite{SharanN08}. 

The approximation graph can be used as input to any of the static graph summarization algorithms presented in Section~\ref{sec:static}. However, this approach has the same shortcoming as the straightforward approach---namely, it depends on the time granularity of the input graph sequence. Probabilistic relational models (PRM) and relational Markov decision processes (RMDP, which are a sequence of PRMs forming a chain that follows a first-order Markov assumption) have also been used to model dynamic graphs~\cite{GuestrinKGK03}, but they cannot model time-varying edges and treat them as fixed over time.

\subsection{\bf Grouping-based methods}
\new{\textit{Grouping-based summarization approaches recursively aggregate nodes and timesteps to reduce the size of large-scale dynamic networks.}}

\new{\netcondense~\citet{adhikari2017condensing} is a node-grouping approach that maintains specific properties of the original time-varying graph, like diffusive properties important in marketing and influence dynamics, governed by its maximum eigenvalue. 
In this context, given a dynamic network of $T$ snapshots and an epidemiology model, the goal is to find a reduced network series with few groups of nodes (supernodes) and groups of timesteps so that the change in its maximum eigenvalue is minimized. 
In its general form, this problem is intractable, but it can be transformed into an equivalent static-graph problem with a well-conditioned, flattened network whose eigenvalue is easy to compute and has similar diffusive properties as the original dynamic network. 
This observation allows solving the dynamic problem with an algorithm similar to \coarsenet~\cite{purohit2014fast}  (Section~\ref{sec:static}). 
In this case, after flattening the dynamic network, \netcondense repeatedly merges adjacent node pairs and adjacent time pairs, evaluating the change in \newnew{the flattened network's} maximum eigenvalue. 
The changes are sorted in increasing order and the best node-/time-pairs are merged until the user-specified network size is achieved. 
\netcondense uses transformations and approximations to achieve sub-quadratic running time and linear space complexity.} 

\new{In many applications such as network monitoring and urban planning, network edges are observed sequentially. Traditional sketching techniques \cite{ZhaoAW11,CormodeGM05}
usually maintain only frequency counts, ``dropping'' the information of the graphical structure, although the goal in summarization is to both construct a summarized \emph{graph} in linear time and to support edge updates in constant time. 
To this end, TCM~\cite{tang2016graph} approximates a variety of graph queries by creating and querying $d$ graph sketches, and returning the minimum answer. Each graph sketch $i$  is created by mapping the original nodes to ``node buckets'' or supernodes via a hash function $h_i$. The edges between supernodes in the graph sketch are superedges corresponding to the sum of the connections between their constituent nodes. The more pairwise independent hash functions (sketches) are used, the lower the probability of hash collisions and thus the more precise are the answers to the queries. 
By maintaining the graphical structure, TCM supports complex analytics over graph streams, such as conditional node queries, aggregated edge weights, aggregated node flows, reachability path queries, aggregate subgraph queries, triangles, and more.}

\subsection{\new{Bit} compression-based methods} 
\textit{The techniques in this category use compression as a means of extracting meaningful patterns from temporal data.} 
This category's only representative is \timecrunch~\cite{ShahKZGF15}, which succinctly describes a large dynamic graph with a set of important temporal structures. Extending \vog~\cite{KoutraKVF14} (Section~\ref{sec:compression}), the authors formalize temporal graph summarization as an information-theoretic optimization problem where the goal is to identify the temporal behaviors of local static structures that collectively minimize the global description length of the dynamic graph. 
A lexicon that describes various types of temporal behavior (flickering, periodic, one-shot) is introduced to augment the vocabulary of static graphs (stars, cliques, bipartite cores, chains). 

\begin{wrapfigure}{r}{0.65\textwidth}
\centering
        \includegraphics[width=0.65\textwidth]{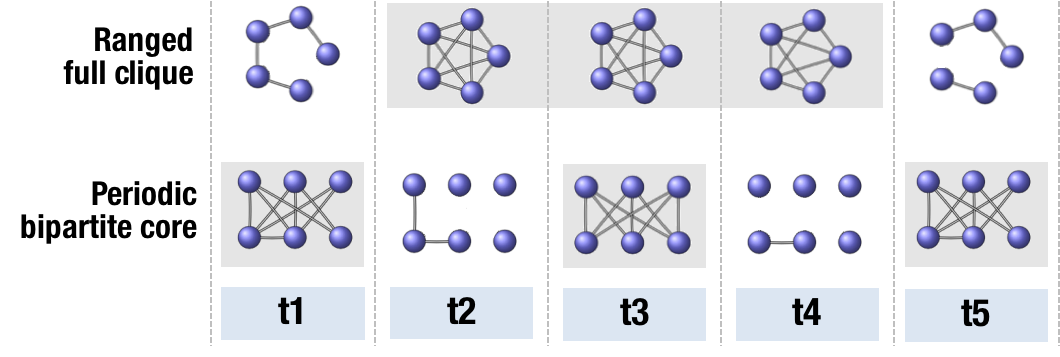} \caption{\newnew{Examples of temporal patterns identified by \timecrunch~\cite{ShahKZGF15}: ranged full clique at times $t2-t4$ and periodic bipartite core every other timestep.} 
        }
        \label{fig:timecrunch}
  \end{wrapfigure}
\timecrunch (i) first identifies static structures in each timestamp, (ii) labels them using the static lexicon, (iii) stitches them together to find temporal structures, (iv) then labels those using the temporal lexicon, and (v) selects for the summary the temporal structures that help minimize the MDL cost of describing the time-evolving graph. Stitching  static structures corresponds to evolution tracking, which is handled via iterative rank-1 singular value decomposition (SVD) to find potentially temporally-coherent structures.
Then, cosine similarity ensures the temporal coherence of the discovered structures. 

  
\subsection{Influence-based methods} 
 
\textit{Influence and diffusion processes are inherently time-evolving. The methods in this category summarize the influence process mainly in dynamic social networks.} 
In Section~\ref{subsec:infl_static} we present two techniques, CSI~\cite{Mehmood2013} and SPINE~\cite{MathioudakisBCGU11}, that summarize social graphs by leveraging information propagation and social influence processes. 
These approaches have a temporal aspect since they are summarizing inherently temporal activities in networks, but they operate on static graphs, where the directed edges capture influence. 

Here we focus on a method that summarizes interestingness-driven diffusion processes in dynamic graphs. The input of OSNet~\cite{QuLJZF14} is a stream of time-ordered interactions, represented as undirected edges between labeled nodes. Its goal is to capture cascades (for example, the spread of news) in a directed graph 
that reveals the flow of dynamics. The output summary consists of subgraphs with ``interesting'' nodes from the original input graph, 
where interestingness is defined as a linear combination of the out-degree of a node (the number of nodes that it infects during the diffusion process), and the maximum ``propagation radius'' (the length of the path from the root of the diffusion process to the node). 
The core technical ideas of OSNet are (i) to construct spreading trees and (ii) to compute the interestingness of a summary via its entropy and a threshold that can lead to fast convergence. OSNet outperforms static-based summarization techniques \cite{ToivonenZHH11,NavlakhaRS08} that give a summary per timestamp, since they are not suited for capturing temporal dynamics; the former depends on user-defined parameters, and the latter gives summaries with many disconnected cliques.

Relatedly, \citet{LinSK08} focus on understanding a social group's collective activity over time. 
To this end, the authors extract activity themes over time using non-negative matrix factorization on a multi-graph (user-photo, user-comment, photo-tag, and comment-tag graphs) in order to obtain latent spaces for users and concepts. 
The top $k$ users and terms in the latent space define the ``important'' actions, which correspond to activity themes. Evolution of themes over time is tracked by applying cosine similarity between their corresponding latent spaces, similar to the evolution tracking component of \timecrunch~\cite{ShahKZGF15}, which also uses cosine similarity to ensure temporal coherence.  
\citet{LinSK08} visualize the themes as bubbles connected by edges, each of which has a length inversely proportional to the similarity of the themes. 



\vspace{0.1cm}
\noindent \textbf{Connections to graph clustering, sparsification, and compression.} As with static graphs, techniques such as clustering, sparsification and compression are related to summarization methods for dynamic graphs. 
Some clustering methods extend heuristics that have been used for static graphs, such as modularity~\cite{GorkeMSW10} or minimum-cut trees~\cite{SahaM07}, and others introduce definitions specific to the temporal domain~\cite{conf/icdm/TantipathananandhB11}. As discussed in Section~\ref{sec:static}, graph sketches~\cite{AhnGM12,Liberty13} summarize large amounts of data by applying linear projections. The property of linearity is fundamental, as it makes sketches applicable to the analysis of streaming graphs in centralized or even distributed settings, where they are partitioned in multiple servers with MapReduce~\cite{Dean04MapReduce}. One-pass and other efficient streaming algorithms with their theoretical analysis are given in \cite{AhnGM12}.

Work on compressing dynamic graphs for storage includes lossy compression of time-evolving graphs~\cite{HeneckaR15}, and encoding of dynamic, weighted graphs as three-dimensional arrays (tensor) by reducing heterogeneity and guaranteeing compression error within bounds~\cite{LiuKCBLPK12}. The latter is based on hierarchical clusters of edge weights  and graph compression using run-length encoding,
traversing first the tensor's time dimension  and second the tensor's vertex dimensions. 
This method maintains the connectivity of the graph as defined by the average shortest paths over all pairs of connected nodes.
Thus, it handles related queries with good approximations.  

\section{Graph Summarization in Real-world Applications}
\label{sec:application}
As we mention in the introduction, summarization helps mitigate information overload. 
In this section, we discuss real-world applications of graph summarization, which are myriad and relevant in many domains.

\new{\subsection{Summarization for query handling and efficiency}}
Graph summarization can greatly improve query execution and efficiency across different graph-specific queries.
\new{Such queries may seek node-related information like degree, PageRank, or participating triangles, or look to identify or match subgraphs within a larger graph.
Table~\ref{tab:query-efficiency} outlines several types of queries used to evaluate graph summarization methods}.
\begin{table}[h!]
\centering
\caption{\newnew{Qualitative comparison of summarization approaches for query handling.}}
\label{tab:query-efficiency}
\resizebox{\textwidth}{!}{
\begin{tabular}{l || ccccccc ||  c } \toprule
 \multirow{7}{*}{\textbf{Method}} &  \multicolumn{7}{c||}{\textbf{Query Type}} &  \multirow{7}{*}{\textbf{Graph Type}} \\
  & \rotatebox[origin=c]{90}{Star} & \rotatebox[origin=c]{90}{Neighborhood} & \rotatebox[origin=c]{90}{Degree} & \rotatebox[origin=c]{90}{Triangles} & \rotatebox[origin=c]{90}{Patt.\ Matching$\;$} & \rotatebox[origin=c]{90}{Reachability} & \rotatebox[origin=c]{90}{PageRank} & \\ 
\midrule 
Graph dedensification~\cite{maccioni2016dedensification} (\S\ref{sec:static})  & \ding{51} & & & & \ding{51} & & &  social, web\\ \hline
$l_p$-reconstr. Error~\cite{riondato2014graph} (\S\ref{sec:static})  & & \ding{51} & \ding{51} & \ding{51} & & & &  social \\ \hline
Query-pres.~\cite{fan2012query} (\S\ref{sec:attribute})  & & & & & \ding{51} & \ding{51} & &  citation, social, web\\ \hline
Neighbor friendly compr.~\cite{MaserratP10} (\S\ref{sec:static})  & & \ding{51} & & & & & & social\\ \hline
GraSS~\cite{lefevre2010grass} (\S\ref{sec:static}) & & & \ding{51} & & & & \ding{51}  & co-authorship, wiki\\ \hline
S-Node~\cite{raghavan2003representing} (\S\ref{sec:attribute})  & & \ding{51} & & & & & \ding{51}  & collaboration, web\\ \hline
$d$-summaries~\cite{song2016knowledgegraphs} (\S\ref{sec:attribute}) & \ding{51}  & &  & & \ding{51} & &   & knowledge graph\\ \hline
\bottomrule
\end{tabular}
}
\end{table}

\new{Pattern-matching queries are extremely common on graph databases. 
For example, across star queries involving nodes of different degrees, graph dedensification~\cite{maccioni2016dedensification} improves query efficiency as the size of the queried graph increases, \newnew{yielding the best improvements (up to $10\times$ speedup)  for queries involving}  
only high-degree nodes (Section~\ref{sec:static}). }
\newnew{\citet{fan2012query} propose an attributed graph compression method and query transformation scheme for lossless pattern matching queries, achieving a compression rate up to 92\% and runtime reduction up to 70\% (Section~\ref{sec:attribute}).} 
\new{From a systems point of view, ~\citet{goasdoue2015query} propose query-oriented graph summarization on Resource Description Framework (RDF) graphs, which are the standard model for W3C web resources.} \newnew{Many methods for pattern matching queries also exist outside graph summarization in databases and graph analytics~\cite{TongFGR07,TianP08,FanWW13,PientaTTC14}, but these are beyond our survey's scope.} 

\new{Summarization has been shown to improve query efficiency in diverse domains.
For example, a typical query on a social network graph might ask whether an edge exists between two nodes, or more generally whether there exists a path between two nodes.
Such a query can be answered on a space-efficient summary of an expected adjacency matrix~\cite{riondato2014graph}. \newnew{This approach constructs the graph summary up to 12500$\times$ faster than its baseline GraSS~\cite{lefevre2010grass}, and also achieves lower average query error. }
Another application is on Web graphs~\cite{raghavan2003representing}:
here, the S-Node representation (Section~\ref{sec:attribute}) outperforms other representation schemes by an order of magnitude on complex web navigation queries by loading only a relatively small number of intranode and superedge graphs and avoiding disk I/Os when possible, \newnew{leading  to 75-90\% reduction in navigation time compared to baselines.}
}
\newnew{A final application is knowledge graphs~\cite{song2016knowledgegraphs}, which lead to up to $40\times$ speedup over an optimized frequent subgraph mining algorithm on generated knowledge graphs for a variety of subgraph queries.}

\subsection{Summarization for visualization and pattern discovery}
\label{sec:app_viz}

 \begin{wrapfigure}{R}{0.32\textwidth}
\vspace{-0.45cm}
\centering
\includegraphics[width=0.3\textwidth]{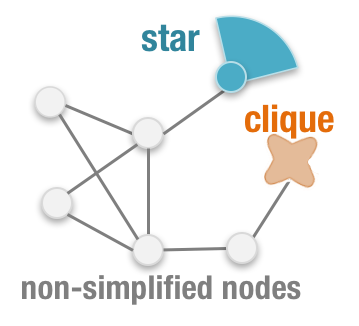}
\caption{\newnew{Example of simplified network with one clique motif glyph and one star motif glyph~\cite{DunneS13}.}
}
\label{fig:motif}
\end{wrapfigure}

Summarization can enable visualization of data too large to load, display, and interactively explore in original raw format. 
For example, ~\citet{shen2006visual} apply OntoVis (Section ~\ref{sec:simplification}) on a large heterogeneous movie network consisting of 8 node types (person, movie, role etc.) with 35,000 nodes and 108,000 links: though relatively small, this graph is still too dense to fit on a desktop screen. To investigate the relationships between persons and roles, the authors visually observe \newnew{the summarized network} and identify a role-actor relationship where a good actor should be able to play different roles (for example, actors like Woody Allen and Sandra Bullock play three different types of roles). 
 Other works that perform visualization on top of summarization include \vog~\cite{koutra:14:vog}, which visualizes structures of specific types (e.g. cliques, bipartite cores), and Motif Simplification~\cite{DunneS13}, which visualizes simplified networks of up to 8,000 nodes with glyphs, a toy example of which is given in Figure~\ref{fig:motif}. 

\newnew{Summarization also supports pattern discovery by maintaining ``interesting'' or ``salient'' patterns.}
Consider the \Wikipedia-\Wikipediacontro dataset, in which nodes are Wikipedia contributors and edges connect users who edit the same part of the article.  ~\citet{koutra:14:vog} apply \vog on this graph to extract the 10 most informative structures, obtaining 8 stars and 2 bipartite subgraphs. 
The centers of the stars correspond to admins or heavily active contributors.
The bipartite cores correspond to edit wars between groups of users, like vandals and responsible editors, on a controversial topic.

SUBDUE~\cite{cook:94:subdue}, one of the most famous frequent-pattern mining methods, is applied in areas as diverse as chemical compound analysis, scene analysis, and CAD circuit design analysis. For example, SUBDUE discovers substructures in chemical compound graphs where atoms are vertices and edges are bonds, in particular discovering the building-block components that are heavily used, such as isoprene units for rubber compounds.

\subsection{Summarization for influence extraction}
Influence analysis is a long-standing research focus and objective of graph mining.
In graph summarization, ~\citet{li2009egocentric} use egocentric abstraction to extract influence from a simulated heterogeneous crime dataset with nodes as gangs and edges as gang relations. In a user study, they demonstrate that the graph abstraction leads to more accurate, efficient, and confident identification of high-level crime-committing gangs. Furthermore, it is demonstrated that each abstraction view captures different parts of key criminal evidence to some extent: for example,  ``the gang has hired a middleman intending to commit a crime''.

Another example is  \coarsenet~\cite{purohit2014fast}: applied on cascade network Flixster of 56,000 nodes and 560,000 edges, it is demonstrated that  a large fraction of movies propagate in a small number of groups with a multi-modal distribution, suggesting movies have multiple scales of spread. Finally, \citet{Mehmood2013} use community-level social influence analysis on Yahoo! and Twitter graphs to observe almost no correlation between influence and link probabilities. 
In other words, it is demonstrated that influence relationships do not in general exhibit any clear structure. 
Even dense communities do not necessarily exhibit strong internal influence. 

\section{Conclusion}
\label{sec:conclusion}
In this survey we present the state-of-the-art in graph summarization. 
Distinguishing between types of input graph and core summarization techniques, we propose a taxonomy to categorize existing graph summarization algorithms. 
We introduce the key details of each algorithm and explore relations between relevant works and methods, also providing examples of real-life applications for each algorithm category. 
Here, we point readers to important open problems in the field.

\subsection{Open Research Problems}

While graph summarization research is advancing, the field is still relatively new and underexplored.
First, further work is to be done in \emph{handling diverse input data types}.
There does not yet exist work on \emph{summarizing temporal graphs with side information}, even though many real-world networks, like social networks, can easily (and perhaps most accurately) be modeled as temporal attributed graphs. 
Even beyond the temporal aspect, other static graph types have yet to be addressed. An example is the \emph{multi-layer graph}, which is an important model for Web graphs~\cite{laura2002multi}; another is the \emph{multiview graph}, which can be ``viewed'' from its different edge types. For instance, a Twitter graph could comprise separate adjacency matrices for follows, retweets, and messages.
As data become increasingly richer, methods will need to handle graphs that comprise multiple views or incorporate other types of data, like time series associated with network nodes.

Another area of improvement is \emph{standardizing, generalizing, or extending} algorithmic and evaluation techniques.
For example, numerous methods tailored toward query efficiency on graph summaries exist, but they either perform approximate queries or are limited to very specific exact queries.
Further work should address lossless compression with \textit{general-purpose queries}.
Another example is labeled graphs: existing methods group nodes with cohesive attribute values, but in some applications \emph{heterogeneous clusters} are crucial. For example, such clusters could facilitate anomaly detection or else refer to groups with desired diversity, as in an academic or professional setting.
In terms of evaluation, current measures are also usually highly application-specific.
Compression-based methods are evaluated on compression quality; query-oriented methods are evaluated on query latency; and so on. Some common evaluation metrics can make comparison of new and established approaches easier: for example, metrics that evaluate supergraphs on sparsity, least information loss, and ease of visualization.

Finally, a promising new direction is graph summarization using \textit{deep node representations} learned automatically from the context encoded in the graph. Node representation learning 
has attracted significant interest in recent  years~\cite{deepwalk,node2vec,WangCZ16,TangQWZYM15}. Given the existence of summarization methods using latent node representations (e.g., via factorization) or manually selected node/egonet/$k$-hop neighborhood features, as well as the recent successes of deep learning, deep node representations for summarization naturally seem promising.

Overall, summarization methods are becoming increasingly important and useful as the volume of available interconnected data rapidly grows.
While we overview several formulations of graph summarization already studied, we conclude by noting that many promising directions in the field remain unexplored and thus full of potential for impact.




\bibliographystyle{ACM-Reference-Format-Journals}
\bibliography{abbrev,all,survey}

\end{document}